\documentclass[10pt,a4paper]{article}
\usepackage[utf8]{inputenc}
\usepackage{amsmath}
\usepackage{amsfonts}
\usepackage{xcolor}
\usepackage{amssymb}
\usepackage{verbatim}

\usepackage{authblk}
\usepackage{graphicx}
\usepackage[left=3cm,right=3cm,top=2.5cm,bottom=2cm]{geometry}
\usepackage{caption}
\usepackage{subcaption}

\date{}

\title{Nonlinear spatial evolution of degenerate quartets of water waves}

\newcommand{\cga}{c_{g,a}}
\newcommand{\cgb}{c_{g,b}}
\newcommand{\cgc}{c_{g,c}}
\newcommand{\B}[1]{|B_{#1}|}

\author[1]{Conor Heffernan}

\author[2,3]{Amin Chabchoub}

\author[1]{Raphael Stuhlmeier}

\affil[1]{{School of Engineering, Computing \& Mathematics},%Department and Organization
            {University of Plymouth}, 
            {Plymouth},
            {PL4 8AA}, 
            {United Kingdom}}

\affil[2]{{Hakubi Center for Advanced Research},
             {Kyoto University},
             {Kyoto},
             {606-8501},
             {Japan}}
             
\affil[3]{{Disaster Prevention Research Insitute},
             {Kyoto University},
             {Uji},
             {611-0011},
             {Japan}}

\begin{document}

\maketitle


\begin{abstract}
In this manuscript we investigate the Benjamin-Feir (or modulation) instability for the spatial evolution of water waves from the perspective of the discrete, spatial Zakharov equation, which captures cubically nonlinear {and resonant wave} interactions in deep water without restrictions on spectral bandwidth. Spatial evolution, with measurements at discrete locations, is pertinent for laboratory hydrodynamic experiments, such as in wave flumes, which rely on time-series measurements at a series of fixed gauges installed along the facility. This setting is likewise appropriate for experiments in electromagnetic and  plasma waves. Through a reformulation of the problem for a degenerate quartet, we bring to bear techniques of phase-plane analysis which elucidate the full dynamics without recourse to linear stability analysis. In particular we find hitherto unexplored breather solutions and discuss the optimal transfer of energy from carrier to sidebands. Finally, we discuss the observability of such discrete solutions in light of numerical simulations.
\end{abstract}

\section{Introduction}

The Benjamin-Feir (or modulation) instability of waves in deep water is one of the most prominent discoveries of nonlinear science during the 20th century. The fact that monochromatic waves distort while propagating in the laboratory, and that these difficulties can be attributed to fundamental energy transfers which impact our ability to forecast waves or understand extreme events continues to be a source of fascination to this day.

The potential flow problem with free surface which describes water wave propagation is nonlinear due to the surface boundary conditions. To deal with this formidable problem, the investigation of the instability of a monochromatic wave train by Benjamin \& Feir \cite{Benjamin1967a} employed perturbation theory in the spirit of G.~G.~Stokes, who pioneered its use in hydrodynamics more than a century earlier. When the problem is linearised, periodic, traveling waves consisting of a single Fourier harmonic are easily found. From the second order in the perturbation expansion these waves change shape due to the addition of bound harmonic terms, and at third order undergo a first dispersion correction, such that their frequency depends on their amplitude. These changes were known to Stokes by the mid 19th century. 

What Stokes could not anticipate was how the dynamics would change if more than one Fourier harmonic were present in the lowest order solution. Indeed, the fact that the onerous perturbation expansions might yield something worthwhile became apparent only with Phillips' \cite{Phillips1960} discovery of a resonant mechanism for energy exchange among water waves at third order. This gave the impetus to explore the problem more deeply, introducing initially small superharmonic and subharmonic perturbations into the water wave problem, and culminating in the work of Benjamin \& Feir. Simultaneously with these developments, efforts were underway to develop compact model equations for the nonlinear evolution of perturbed wave trains.

In water waves this resulted in the groundbreaking 1968 paper of Vladimir Zakharov \cite{Zakharov1968}, wherein a Hamiltonian formulation of the water wave problem was determined for the first time. In the same work this novel formulation was used to derive a nonlinear Schr\"{o}dinger equation (NLS) in the context of hydrodynamics, which was then employed to analyse the linear stability of uniform wave trains. The NLS is in fact a type of universal equation for the slow variation of wave envelopes, as previously shown by Benney \& Newell \cite{Benney1967}. A linear stability analysis based on such a model equation had previously been employed by Bogoliubov \cite{Bogoliubov1947} in understanding the elementary excitations of a Bose-Einstein condensate, and by Bespalov \& Talanov \cite{Bespalov1966} for nonlinear electromagnetic waves.

These analyses, starting from the Gross-Pitaevski or nonlinear Schr\"odinger equations, allow for the study of small perturbations -- modulations -- of a plane wave solution. This is effectively a Fourier mode truncation, whereby the partial differential equation is reduced to a system of coupled ordinary differential equations which are linearised to obtain a threshold for stability. The initial exponential growth soon renders the linearisation invalid, and understanding the subsequent behaviour requires new methods. One natural avenue of progress was numerical investigation, although this comes with its own pitfalls, see, for example, Ablowitz \& Herbst \cite{Ablowitz1990}. As numerical solutions require the fine-tuning of initial conditions and inevitably involve computations with finite accuracy, the complementary route of seeking exact solutions came to play a significant role.
From the point of view of the nonlinear Schr\"odinger equation, the most remarkable exact solutions are the breathers found by Kuznetsov \cite{Kuznetsov1977} and Ma \cite{Ma1979}, Akhmediev et al \cite{Akhmediev1987} and Peregrine \cite{Peregrine1983a}. These solutions represent the reversible amplification of disturbances from a background state into a spatially, temporally or spatiotemporally-localised coherent structure, and have been the subject of intense interest since their discovery.

Another line of inquiry, led initially by the Fluid Mechanics Department at TRW Defense and Space System and collaborators, and reviewed in Yuen \& Lake \cite{Yuen1982}, sought to explore the modulation instability using relaxed assumptions. In particular, an intermediate result in Zakharov's seminal 1968 paper \cite{Zakharov1968} -- namely a reduced form of the Hamiltonian equation which retains only resonant contributions -- provided a window into the instability without some of the narrow-bandwidth restrictions imposed by the NLS formulation. Employing this equation, which became known as the Zakharov equation, Crawford et al \cite{Crawford1981} gave improved linear stability bounds, and contributed to a profusion of interest in modulation instability with broader bandwidth \cite{Dysthe1979,TRULSEN1996}.
This reduced Hamiltonian equation has, in fact, a rather generic form \cite{Zakharov1992a}: for a dispersion law which permits four wave resonances but not three wave resonances, the particular physics of the problem are contained only in the corresponding integral kernel. In the context of water waves, the correct Hamiltonian form of this kernel was given by Krasitskii \cite{Krasitskii1994}. Analogous reduced Hamiltonians can be found, for example, for Langmuir waves in plasma or optical waves described by the Maxwell equation \cite{Onorato2016}. 

Historically, most approaches to the problem of modulation instability and its consequences focus on  equations written in terms of temporal evolution. While this is natural mathematically, as it is common to initiate the dynamics from initial conditions, it does not correspond to the typical experimental set-up in hydrodynamics, where times-series of surface wave evolution are measured along the tank by means of wave gauges placed at fixed spatial locations, i.e., the wave dynamics are initiated through boundary conditions. This calls for a suitable spatial evolution equation, either in the form of the NLS \cite{Chabchoub2016a} or a spatial Zakharov equation \cite{Shemer2001}. The latter equation, derived by Shemer and co-workers in the early 2000s, has been used for a small handful of studies (e.g.\ \cite{Shemer2002,Galvagno2021}), but its consequences for the Benjamin-Feir instability have barely been explored (one notable exception is Shemer \& Chernyshova \cite{Shemer2017}).

Our objective is to employ the spatial Zakharov formulation, which arises from the cubically nonlinear problem but otherwise makes no assumptions about spectral bandwidth, to understand the entire spatial evolution of modulation instability. Exactly as in studies of the instability threshold using NLS we shall begin by truncating our system to three interacting Fourier modes. In contrast to the classical approach, we find a subsequent linearisation to be superfluous: the resulting system can be recast as a planar Hamiltonian dynamical system, whose dynamics can be analysed by studying fixed points, separatrices, and bifurcations.  This dynamical system naturally encompasses the bi-modal spectrum \cite{Shemer2017}, which is in fact the natural counterpoint of the classical Benjamin-Feir instability. 

Our approach provides analytical insight into new solutions {and associated novel physics} of this spatial equation, including an analogues of the Kuznetsov-Ma breather solution. Moreover it shows a theoretical route towards optimal conversion of energy from a monochromatic to a bichromatic sea, or vice versa. We explore the stability of this optimal depletion solution to higher harmonics both analytically and numerically, in a complement to the phase plane analysis for the interacting degenerate quartet.

We now give an outline of the subsequent sections of this paper. In Section \ref{sec: Fundamentals}, we provide a description of the fundamentals of the Zakharov equation. We show how it is discretised to a finite number of interacting waves which lays the groundwork for our analysis, before ending with a discussion of special cases, which admit closed form solutions. In Section \ref{sec:Reformulation of discrete ZE}, we build on these fundamentals discussed in Section \ref{sec: Fundamentals} as we consider the degenerate quartet case of three distinct frequencies and reduce the Zakharov equation to a discrete set of ODEs for the amplitude and phase of the interacting waves. By considering certain conserved quantities, we can reduce the dynamics to a two dimensional dynamical system. 

In Section \ref{sec: Dynamics of interacting waves} we discuss the various phenomena that result from three interacting waves. We first discuss the periodic behaviour which represents the generic, recurrent evolution of the Benjamin-Feir instability -- also called Fermi-Pasta-Ulam-Tsingou recurrence. We then analyse the breather solutions present in our system, arising from both monochromatic and bichromatic background states. We also discuss the issue of maximum energy transfer away from the carrier, which is found in a particular breather solution. In Section \ref{sec:Observability of discrete wave interactions}, we discuss the observability of these wave phenomena when additional Fourier modes are present, and consider the implications for wave flume experiments. Finally, in Section \ref{sec: Discussion}, we conclude with a summary of our work as well as suggestions for some avenues of future study.

\section{Fundamentals}
\label{sec: Fundamentals}

Our starting point will be the spatial Zakharov equation developed in the early 2000s by Shemer et al \cite{Shemer2001}. This is a modification of the temporal evolution equation for {nonlinear waves}  derived by Zakharov \cite{Zakharov1968}, and in Hamiltonian form by Krasitskii \cite{Krasitskii1994}. The spatial Zakharov equation has the form
\begin{align}\nonumber
i c_g \frac{\partial B(x,\omega)}{\partial x} = &\iiint T(k,k_1,k_{2},k_{3}) B^*(x,\omega_1) B(x,\omega_2) B(x,\omega_3) \\
&\cdot  \exp(-i(k+k_1-k_2-k_3)x) \delta(\omega+\omega_1 - \omega_2 - \omega_3) d \omega_1 d \omega_2 d \omega_3.
\label{eq:Spatial ZE}
\end{align}
where $c_g$ denotes the deep-water linear group velocity, $k_i$ is a wavenumber, $\omega_i = \omega(k_i) = \sqrt(g k_i)$ the linear dispersion relation in deep water, and $*$ denotes the complex conjugate. By $\delta$ we denote the Dirac delta distribution and each integral is taken over the real line. The complex amplitudes $B$ are related to the Fourier transforms of the free-surface elevation $\eta$ and the potential at the free surface, which may be recovered from this formulation (see below). Full expressions for the kernel $T(k,k_1,k_2,k_3)$ are available in \cite{Krasitskii1994}. We note only that this kernel has the following symmetries:
\[ T(i,j,k,l) = T(j,i,k,l) = T(i,j,l,k) = T(k,l,i,j). \]
It is common to write the wavenumber-dependence of the kernels as a subscript, and we shall employ the abbreviation $T_{jlmn}$ for the kernel $T(k_j,k_l,k_m,k_n).$
Owing to the symmetries we will also denote $T_{jjjj}=T_j$ and $T_{ijij}=T_{ij}$ without risk of confusion.

This equation can be discretised as follows:
\begin{equation}
\label{eq:Spatial ZE Discr}
i c_{g,j} \frac{d B_j(x)}{dx} = \sum_{l,m,n} T_{jlmn} B_l^* B_m B_n \exp(-i(k_j + k_l - k_m - k_n)x) \delta(\omega_j + \omega_l - \omega_m - \omega_n),
\end{equation}
where $B_i=B(\omega_i,x),$ and we denote by $\Delta_{jl}^{mn}=k_j + k_l - k_m - k_n$ the wavenumber detuning, which is of order $O(\epsilon^2)$ for $\epsilon$ a characteristic wave steepness \cite{Shemer2001}. 

The relationship between the complex amplitudes and the free surface elevation is given (to lowest order) by 
\begin{equation} \label{eq:B-to-eta}
\eta(x,t) = \frac{1}{2\pi} \int_{-\infty}^{\infty} \left( \frac{\omega}{2g} \right)^{1/2} \left[ B(x,\omega) \exp(i(k(\omega)x-\omega t)) + \text{c.c.} \right] d \omega.
\end{equation}
Here ``c.c.'' stands for the complex conjugate of the preceding expression. More detail on the background and derivation of this equation can be found in the recent review by Stuhlmeier \cite{Stuhlmeier2024}.

\subsection{Simple solutions of the discrete ZE}
\label{sec:Simple Solutions}

The spatial Zakharov equation \eqref{eq:Spatial ZE} can be explicitly solved in some special cases, two of which are of particular interest for a study of the Benjamin-Feir instability. We first consider the case of a single wave $\omega_0,$ such that the spatial Zakharov equation becomes
\begin{equation}
i c_{g,0} \frac{d B_0(x)}{dx} = T_{0} |B_0(x)|^2 B_0(x),
\end{equation}
which admits the constant amplitude solution
\begin{equation} \label{eq: SZE Single Mode Solution}
B_0(x)= A_0 e^{-i A_0^2 T_{0} x/c_{g,0}}. 
\end{equation}
This is a monochromatic wave field with a nonlinear correction to the wavenumber, corresponding to the free mode part of the well-known third-order Stokes' wave solution. Inserting into \eqref{eq:B-to-eta} and using $T_{j} = \frac{k_j^3}{4 \pi^2}$ it can be written as

\begin{equation} 
\label{eq:Spatial Stokes' solution} \eta(x,t) = a_0 \cos (k_0[1 - a_0^2 k_0^2]x-\omega_0 t). 
\end{equation}
where we have normalised the constant amplitude via
\[  A_0 = \pi a_0 \left( \frac{2g}{\omega_0} \right)^{1/2}.\]

The second simple case consists of two waves $\omega_a$ and $\omega_b,$ resulting in a system of two equations
\begin{subequations}
\label{eq:Two mode discretisation}
\begin{align}
i c_{g,a} \frac{d B_a}{dx} &= T_{a} |B_a|^2 B_a + 2 T_{ab} |B_b|^2 B_a,\\
i c_{g,b} \frac{d B_b}{dx} &= T_{b} |B_b|^2 B_b + 2 T_{ab} |B_a|^2 B_b,
\end{align}
\end{subequations}
with solution 
\begin{subequations}
\label{eq:Two mode solution}
\begin{align}
B_a(x) &= A_a \exp ( - i (T_a A_a^2 + 2 T_{ab} A_b^2)x/c_{g,a}),\\
B_b(x) &= A_b \exp ( - i (T_b A_b^2 + 2 T_{ab} A_a^2)x/c_{g,b}),
\end{align}
\end{subequations}
for $A_a$ and $A_b$ two constant amplitudes. The free surface is then a bichromatic (sometimes called bimodal) sea-state, written 
\begin{equation} 
\label{eq:Spatial two mode solution} \eta(x,t) = a_a \cos\left(k_a\left[1 - a_a^2 k_a^2 - 2 a_b^2 k_a^{3/2} k_b^{1/2}\right]x - \omega_a t\right) + a_b \cos\left(k_b\left[1 - a_b^2 k_b^2 - 2 a_a^2 k_a^{3/2}k_b^{1/2}\right]x - \omega_b t\right), 
\end{equation}
where we take $k_a < k_b$ to resolve the two-wavenumber kernels as $T_{ab} = \frac{k_a^2 k_b}{4 \pi^2}.$ This is the spatial counterpart of the solution found using perturbation theory by Longuet-Higgins \& Phillips \cite{Longuet-Higgins1962d}.

\section{Reformulation of the discrete ZE}
\label{sec:Reformulation of discrete ZE}

As soon as more than two Fourier modes are involved the equations become more cumbersome. The principal reason is the appearance of nontrivial interactions beyond the symmetric resonances encountered in Section \ref{sec:Simple Solutions}. In particular, three modes may interact to exchange energy if $2\omega_1 = \omega_2 + \omega_3.$ A description of the resulting interaction is significantly simplified by writing the complex amplitudes in terms of magnitude and phase, as has been suggested and carried out by Bretherton \cite{Bretherton1964}, Craik \cite{Craik1986a}, Capellini \& Trillo \cite{Cappellini1991}, Trillo \& Wabnitz \cite{trillo1991dynamics} (in the context of the NLS), and Andrade \& Stuhlmeier \cite{Andrade2023instability,Andrade2023} (for the temporal Zakharov equation).

In the discrete spatial Zakharov equation \eqref{eq:Spatial ZE Discr} we write the complex amplitude $B_j(x)$ as $|B_j|\exp(i \phi_j),$ where both magnitude and phase may depend on $x.$
Separating into real and imaginary parts leads to:
\begin{align} \label{eq:ZE Magnitude Eq}
& c_{g,j} \frac{d |B_j|}{dx} = - \sum_{l,m,n} T_{jlmn} \delta_{jl}^{mn} |B_l||B_m||B_n| \sin(\Delta_{jl}^{mn} x +\theta_{jlmn}),\\ \label{eq:ZE Phase Eq}
& -c_{g,j} |B_j| \frac{d \phi_j}{dx} = \sum_{l,m,n} T_{jlmn} \delta_{jl}^{mn} |B_l||B_m||B_n| \cos(\Delta_{jl}^{mn} x +\theta_{jlmn}), 
\end{align}
 with
\begin{align*}
{\theta_{jlmn} = \phi_j + \phi_l - \phi_m -\phi_n,}
 \end{align*}
We identify  $\Delta_{jl}^{mn}$ as the wavenumber detuning parameter.

\subsection{Reduction to a degenerate quartet}
If we assume that the indices in equations \eqref{eq:ZE Magnitude Eq}--\eqref{eq:ZE Phase Eq} take on values in the set $\{a,b,c\}$ only, with the proviso that \[ 2 \omega_a = \omega_b + \omega_c, \] so that the resonance condition imposed by the Kronecker delta $\delta_{jl}^{mn}$ is fulfilled, we obtain a set of six ODEs:
\begin{subequations}
\begin{align}
    \cga \B{a}' &= -2 T_{aabc} \B{a}\B{b}\B{c} \sin(\Delta_{aa}^{bc}x + \theta_{aabc})\\
    \cgc \B{c}' &=  T_{aabc} \B{a}^2 \B{b} \sin(\Delta_{aa}^{bc}x + \theta_{aabc})\\
    \cgb \B{b}' &=  T_{aabc} \B{a}^2 \B{c} \sin(\Delta_{aa}^{bc}x + \theta_{aabc})\\
    -\cga \B{a} \phi_a' &= \Gamma_a + 2 T_{aabc} \B{a}\B{b}\B{c} \cos(\Delta_{aa}^{bc}x + \theta_{aabc})\\
    -\cgb \B{b} \phi_b' &= \Gamma_b + T_{aabc} \B{a}^2 \B{c} \cos(\Delta_{aa}^{bc}x + \theta_{aabc})\\
    -\cgc \B{c} \phi_c' &= \Gamma_c + T_{aabc} \B{a}^2 \B{b} \cos(\Delta_{aa}^{bc}x + \theta_{aabc})
\end{align}
\end{subequations}
where 
\begin{equation}
    \Gamma_i = \B{i}^3 T_i + 2 \sum_{j \neq i} \B{i} \B{j}^2 T_{ij}.
\end{equation}
As we are restricted to this so-called degenerate quartet we shall henceforth drop the sub and superscripts on the detuning parameter $\Delta_{aa}^{bc}$ where there is no risk of confusion.

A key observation is that the phases $\phi_i$ of the individual modes appear only in the single combination $\Theta = \Delta_{}^{}x + \theta_{aabc},$ which we identify as the combined (or dynamic) phase variable of the problem. We write an evolution equation for this dynamic phase variable \cite{Bustamante2009,Bustamante2009a} as 
\begin{align} \nonumber
    \frac{d \Theta}{dx} &= \frac{d}{dx} \left( \Delta_{}^{}x + \theta_{aabc} \right) = \Delta_{}^{} + 2\phi_a' - \phi_b' - \phi_c' \\
    &= \Delta_{}^{} - \left( \frac{2 \Omega_a}{\cga} - \frac{\Omega_b}{\cgb} - \frac{\Omega_c}{\cgc}  \right) - T_{aabc} \cos(\Theta) \left( \frac{4 \B{b}\B{c}}{\cga} - \frac{\B{c}\B{a}^2}{\cgb \B{b}} - \frac{\B{b}\B{a}^2}{\cgc \B{c}} \right),
\end{align}
where we use $'$ to denote the derivative in $x$ and define 
\begin{equation}
     \Omega_i = \Gamma_i/\B{i} = \B{i}^2 T_i + 2 \sum_{j \neq i} \B{j}^2 T_{ij}.
\end{equation}
Note that this differs from \cite[Eq.\ (2.5)]{Andrade2023} in the signs of all terms except the first. This is due to the temporal Zakharov equation \cite[Eq.\ (2.1)]{Andrade2023} containing a term $\exp(i\Delta_{np}^{qr}t)$ rather than the term $\exp(-i\Delta_{jl}^{mn}x)$ found in \eqref{eq:Spatial ZE Discr}. %This means that the dynamic phase variable defined in \cite{Andrade2023} is $\Delta_{aa}^{bc}t - 2 \phi_a + \phi_b + \phi_c$ (note that notation has been adapted for consistency).

We find that a quantity akin to wave action 
\begin{equation} \label{eq:Cons WA}
     c_{g,a}|B_a|^2 + c_{g,b}|B_b|^2 + c_{g,c}|B_c|^2 = A
\end{equation}
is conserved, as is the difference in side-band magnitudes
\begin{equation} \label{eq:Cons SB-Diff}
    {c_{g,b} |B_b|^2} - {c_{g,c} |B_c|^2} = A \alpha.
\end{equation}
In light of these conserved quantities, it is useful to reformulate our equations in terms of the squared magnitudes $I_i:=\B{i}^2,$ whereupon 
\[ \frac{d}{dx}I_i = 2\frac{d\B{i}}{dx}\B{i} \] and we obtain the three equations 
\begin{align}
     I_a' &= \frac{-4 T_{aabc}}{\cga} I_a \sqrt{I_b I_c}  \sin(\Theta), \\
     I_b' &= \frac{2 T_{aabc}}{\cgb} I_a \sqrt{I_b I_c}  \sin(\Theta),\\
     I_c' &= \frac{2 T_{aabc}}{\cgc} I_a \sqrt{I_b I_c}  \sin(\Theta).
\end{align}

Making the substitution 
\begin{align}
    I_a &= \frac{A}{c_{g,a}}\eta, \label{eq:singlevar1}\\
    I_b &= \frac{A}{2 c_{g,b}} (1-\eta + \alpha), \label{eq:singlevar2}\\
    I_c &= \frac{A}{2 c_{g,c}} (1-\eta - \alpha).\label{eq:singlevar3}
\end{align}
finally reduces the six coupled equations for the magnitudes and phases to a dynamical system involving two parameters $A$ and $\alpha:$
\begin{align}
\frac{d \eta}{dx} &=  -\frac{2 T_{aabc} \eta  A \sqrt{(1-\eta)^2 -\alpha^2} \sin (\Theta )}{c_{g ,a} \sqrt{c_{g ,b} c_{g ,c}} } \\
    \frac{d \Theta}{dx} &= \frac{2 A T_{aabc} (\alpha^2 - 2 \eta^2 + 3\eta - 1)\cos(\Theta)}{\cga \sqrt{\cgb \cgc} \sqrt{(1 - \eta)^2 - \alpha^2}} + A \eta \Xi_1 + A \Xi_0 + \Delta\end{align}
where 
\begin{align}
   & \Xi_1 = -2 \left(\left( {\tilde{T}_a} -2  {\tilde{T}_{ab}} +\frac{1}{4} {\tilde{T}_b} \right) +\left(-2 {\tilde{T}_{ac}} +{\tilde{T}_{bc}} \right) +\frac{{\tilde{T}_c} }{4}\right) \\
   & \Xi_0 = \frac{\left(\left(1+\alpha \right) \left(-4 {\tilde{T}_{ab}} +{\tilde{T}_b} \right) +4 \left( {\tilde{T}_{ac}} \left(-1+\alpha \right)+{\tilde{T}_{bc}} \right) -{\tilde{T}_c}  \left(-1+\alpha \right)\right) }{2 }
\end{align}
and for compactness we write $\tilde{T}_{ij} = T_{ij}/(c_{g,i}c_{g,j})$ and $\tilde{T}_i=T_i/c_{g,i}^2.$

When the side-band energy is equally distributed and $\alpha=0$ this reduces to the simpler system 
\begin{align}
    \frac{d\eta}{dx} &= \beta_2 \eta (\eta-1) \sin(\Theta) \label{eq:dyn1}\\
    \frac{d\Theta}{dx} &=\beta_2 (2\eta - 1) \cos(\Theta) + \beta_1 \eta + \beta_0 \label{eq:dyn2}
\end{align}
where 
\begin{align}
\beta_2 &= A\frac{2 T_{aabc}}{c_{g,a}\sqrt{c_{g,b}c_{g,c}}}, \\
\beta_1 &= A\frac{\left(-{\tilde{T}_b} -4  {\tilde{T}_{bc}} - {\tilde{T}_c} \right) +\left(8 {\tilde{T}_{ab}} +8 {\tilde{T}_{ac}} \right) -4 {\tilde{T}_a} }{2},\\
\beta_0 &= \Delta + A\frac{ \left(-4 {\tilde{T}_{ab}} +{\tilde{T}_b} \right) +\left(-4 {\tilde{T}_{ac}} +4 {\tilde{T}_{bc}} \right) +{\tilde{T}_c} }{2}.
\end{align}

It is possible to integrate the evolution equation for $\eta$ with respect to $\Theta$ and the evolution equation for $\Theta$ with respect to $\eta$ and so obtain a Hamiltonian for the planar system being analysed. We write the Hamiltonian for $\alpha=0$ as 
\begin{equation}
    H(\eta,\Theta) = - \beta_2 \eta(\eta-1)\cos \left(\Theta \right) - \frac{\eta^2}{2} \beta_1 - \beta_0 \eta.
\end{equation}
Despite the significant differences in the spatial and temporal formulations, this transformed, planar Hamiltonian is essentially analogous to that found in the temporal case by Andrade \& Stuhlmeier \cite{Andrade2023}.

\subsection{Phase-plane dynamics and fixed points}

The fact that we have a planar Hamiltonian dynamical system means it is a simple matter to compute the trajectories, which are simply the level lines of the Hamiltonian. The phase space is the truncated cylinder $\mathbf{p}\in C = \{(\Theta,\eta)\in\mathbb{T}\times\mathbb{R} : -\pi \leq \Theta \leq\pi, 0 \leq \eta \leq 1\},$ whose top $\eta=1$ corresponds to monochromatic waves and whose bottom $\eta = 0$ corresponds to bichromatic waves (see Section \ref{sec:Simple Solutions}) as can be seen immediately from \eqref{eq:singlevar1}--\eqref{eq:singlevar3}. Points in the interior correspond to some mixing of three modes. 

The specification of a trajectory requires that we fix the Fourier modes under consideration, which is done by selecting a central (or carrier) frequency $f$ (1/s) or $\omega$ (rad/s) and a mode separation parameter $p$ such that $\omega_a=\omega=2 \pi f,$ $\omega_b = \omega-p$ and $\omega_c = \omega+p.$ In addition we must specify $A$ in some physically meaningful sense. This is simplest if we employ the relations between the complex amplitudes $B$, the energy scale parameter $\eta$, and the free surface elevation $\eta$ in \eqref{eq:B-to-eta}. 

\begin{figure}[ht]
\centering
    \includegraphics[width=\linewidth]{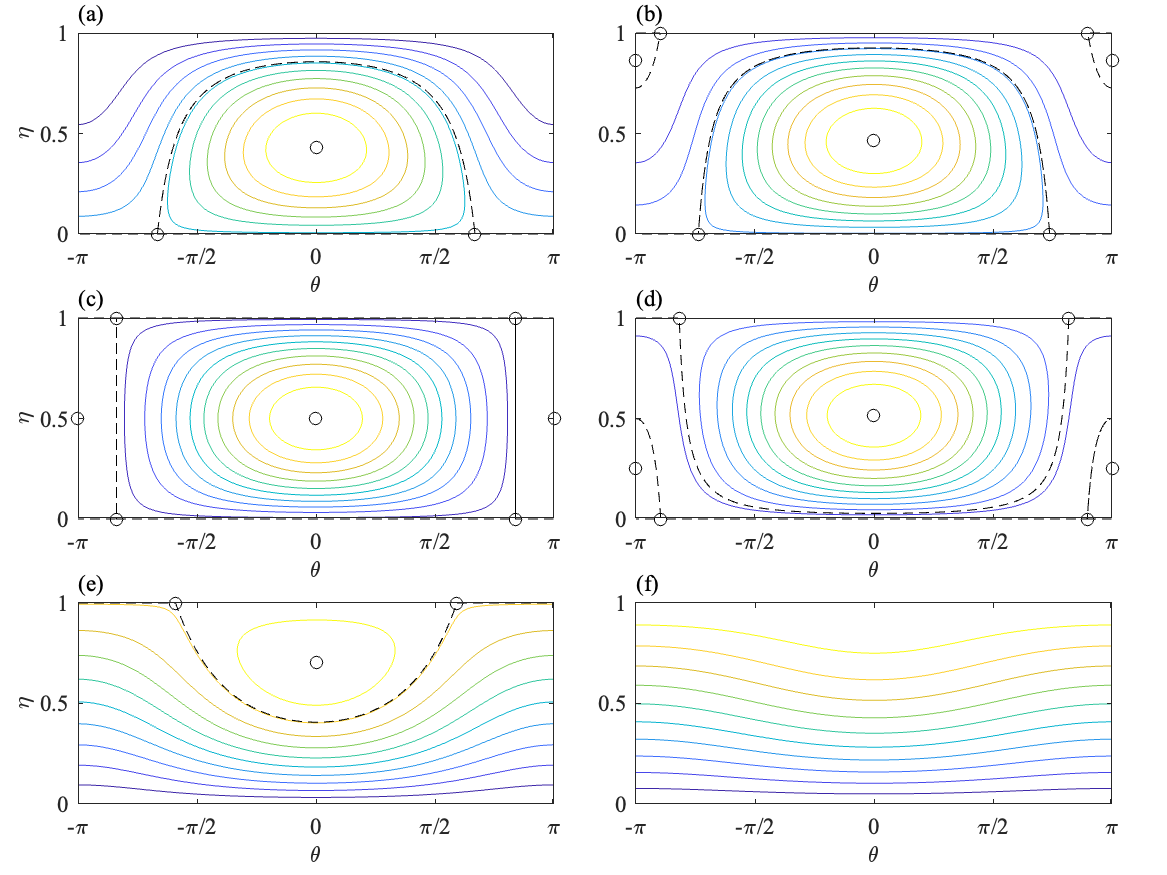}
    \caption{Phase portraits for $(\Theta,\eta)$ for various values of the mode separation parameter $p$ with $f=1$ and $\epsilon=0.2$. Panels (a)--(f) show $p=0, \, 0.6p_c, \, p_c, \, 1.1p_c, \, 2p_c$ and $3p_c$ respectively for $p_c \approx 0.5791.$ Fixed points of the dynamical system are denoted by black circles, separatrices are denoted by dashed curves connecting pairs of fixed points.}
    \label{fig:phase portraits}
\end{figure}

We write this correspondence for a single wave, which we can think of classifying a phase portrait based on the monochromatic carrier \eqref{eq:Spatial Stokes' solution} which forms the top of the phase space. Substituting $\eta=1$ into \eqref{eq:singlevar1}--\eqref{eq:singlevar3} with $\alpha=0$ shows $A=\cga |B_a|^2$ and yields the relation
\begin{equation}
    A = \frac{a_a^2 \pi^2 g}{k_a} = \frac{\epsilon_a^2 \pi^2 g}{k_a^{3}},
\end{equation}
where $a_a$ is the physical amplitude of the carrier wave, and $\epsilon_a = a_a k_a$ is the wave steepness.
The dynamics in the phase space are governed by fixed points and associated separatrices, which are depicted as black circles and dashed curves in Figure \ref{fig:phase portraits}. By setting the right-hand side of \eqref{eq:dyn1}--\eqref{eq:dyn2} equal to zero we can find expressions for the fixed points of the system.

\begin{align}
    \mathbf{p}_1 &= \left(0, \frac{\beta_2 - \beta_0}{2\beta_2 + \beta_1}\right), \label{eq:fp1} \\
    \mathbf{p}_2 &= \left(\pm \pi, \frac{\beta_0 + \beta_2}{2\beta_2 - \beta_1}\right), \label{eq:fp2} \\
    \mathbf{p}_{3,4} &= \left(\Theta_1, 0\right), \label{eq:fp3} \\
    \mathbf{p}_{5,6} &= \left(\Theta_2, 1\right), \label{eq:fp4} 
\end{align}
with $\Theta_1$ and $\Theta_2$ defined as the solutions to the trigonometric equations
\begin{align}
    \cos(\Theta_1) &= \frac{\beta_0}{\beta_2}, \\
    \cos(\Theta_2) &= -\frac{\beta_0+\beta_1}{\beta_2},
\end{align}

We note that these fixed points must be either centres or saddle points. This follows immediately from considering the Jacobian of the system \eqref{eq:dyn1}--\eqref{eq:dyn2}:
\[ J = \begin{pmatrix} H_{\eta \Theta} & H_{\Theta \Theta} \\ -H_{\eta\eta} & - H_{\eta\Theta} \end{pmatrix} =  \begin{pmatrix}
\beta_2(2\eta -1 )\sin(\Theta) & \beta_2\eta (1-\eta)\cos(\Theta) \\
2\beta_2 \cos(\Theta_1) + \beta_1 & \beta_2(1-2\eta)\sin(\Theta) \
\end{pmatrix} \]
which has vanishing trace and determinant
\[ \det(J) = \beta_2 \eta(1-\eta) \cos(\Theta) \left( 2 \beta_2\cos(\Theta) + \beta_1 \right) - (1-2\eta)^2 \beta_2^2 \sin^2(\Theta).\]
We find that the existence of the $\eta=1$ fixed points depends on the condition

\begin{align}
    -1 \leq -\frac{\beta_0 + \beta_1}{\beta_2} \leq 1,
\end{align}
which we may square to give

\begin{align}
    D &= \left({\Delta} + 2 |B_a|^2 \left(\frac{T_{ab}}{c_{g,b}} + \frac{T_{ac}}{c_{g,c}} - \frac{T_{a}}{c_{g,a}}\right)\right)^2 - \frac{4 T_{aabc}^2}{c_{g,b}c_{g,c}}|B_a|^4,\label{eq:linstab}  
\end{align}
such that fixed points exist when $D\geq 0.$ This is exactly the growth rate found from linear stability analysis of the spatial Benjamin-Feir instability (see also \ref{ssec:Linear stability analysis}). Indeed, if fixed points exist at $\eta=1$ the eigenvalues of the Jacobi matrix -- which satisfy $\lambda_{1,2} = \pm \sqrt{-\det(J)}$ -- are positive exactly when $\det(J)$ is negative. It can be readily verified by substitution that at $\eta=1$
\[ \det(J) = \left( \Delta + 2A \left(\tilde{T}_{ab} + \tilde{T}_{ac} - \tilde{T}_a\right) \right)^2 - 4 A^2 \tilde{T}_{aabc}^2. \]

Existence of $\eta=0$ fixed points (and so the instability of the underlying bichromatic sea) depends on the condition 
\[ -1 \leq \frac{\beta_0}{\beta_2} \leq 1,\]
which may be resolved (recalling that here $A = \cgb |B_b|^2 + \cgc |B_c|^2$), after squaring, as
\begin{equation*}
    \left( \Delta + \frac{\cgb |B_b|^2 + \cgc |B_c|^2}{2}\left[ \tilde{T}_c + \tilde{T}_b - 4 \tilde{T}_{ab} + 4 \tilde{T}_{bc} - 4 \tilde{T}_{ac} \right] \right)^2 \leq \frac{4 T_{aabc}^2 (\cgb |B_b|^2 + \cgc |B_c|^2)^2}{\cga^2 \cgb \cgc}. 
\end{equation*}

\begin{figure}[h]
\centering
    \includegraphics[width=0.49\linewidth]{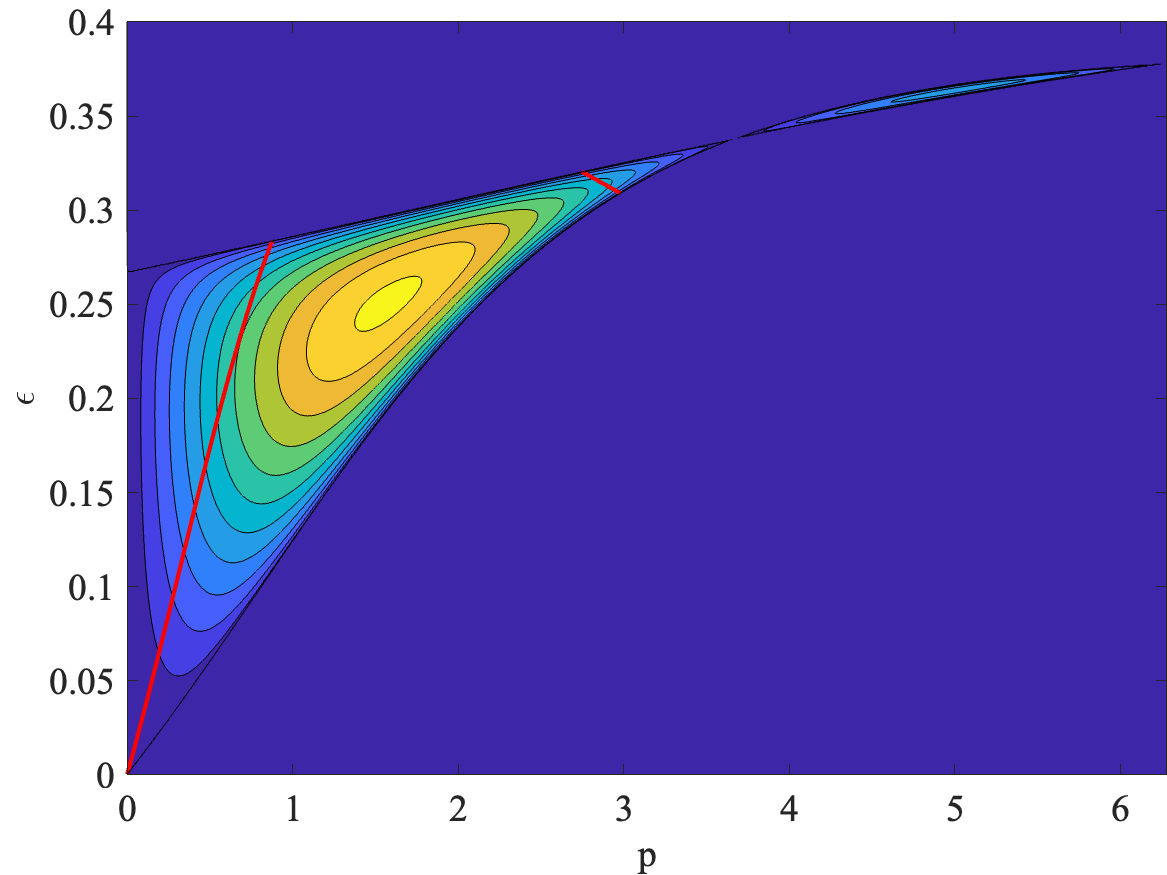}
        \includegraphics[width=0.49\linewidth]{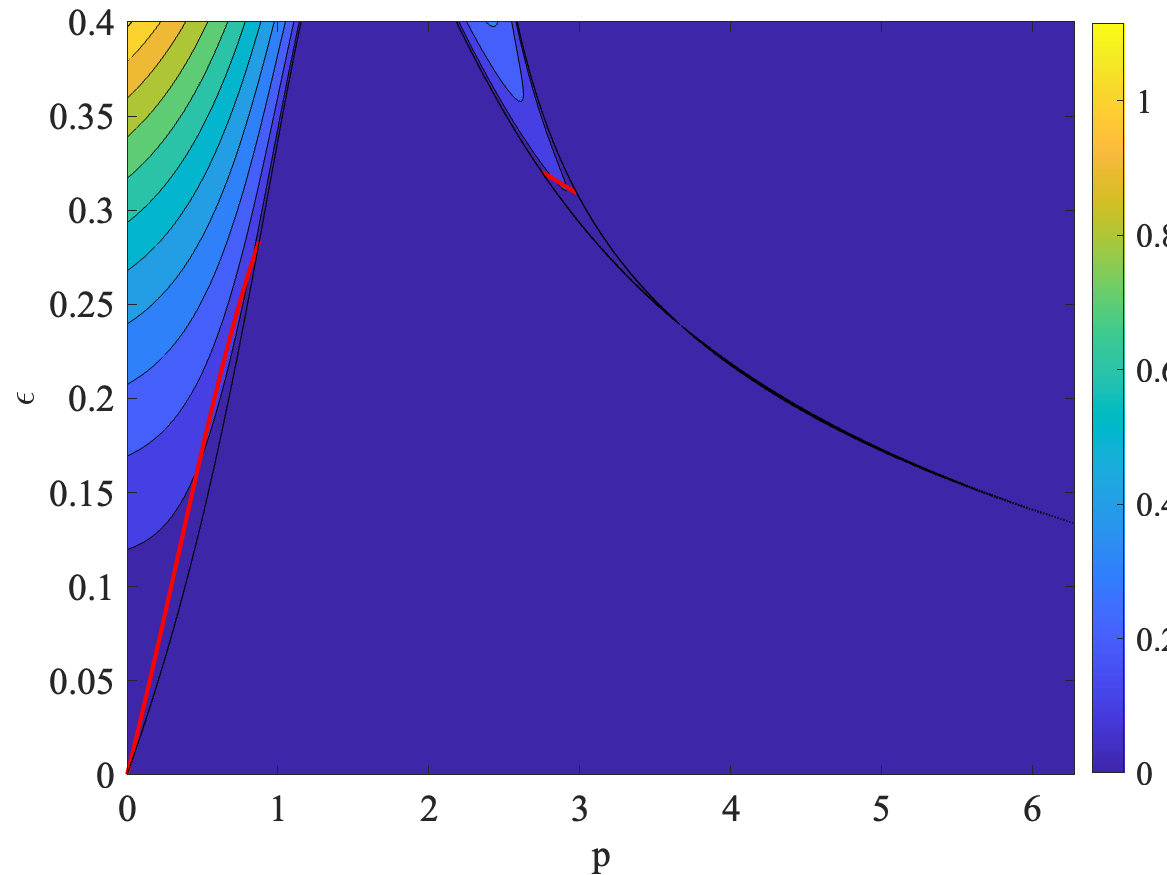}
    \caption{Plot in $\epsilon, \, p$ parameter space of the existence of fixed points (i.e.\ instability) at $\eta=1$ (monochromatic waves, left panel) and $\eta=0$ (bichromatic waves, right panel). Colours denote the growth rate, calculated from the eigenvalues of the Jacobi matrix. The red curve shows the location in $\epsilon, \,p$ space of the maximum depletion of the carrier, associated with the vertical separatrix shown in panel (c) of Figure \ref{fig:phase portraits}.}
    \label{fig:stability regions top/bottom}
\end{figure}

The regions of instability of the monochromatic waves (i.e.\ the spatial Benjamin-Feir instability) and bichromatic waves are shown in Figure \ref{fig:stability regions top/bottom} for fixed carrier frequency $f=1$ Hz. Fixed points appear on the top nullcline $\eta=1$ for values of carrier steepness $\epsilon$ and frequency separation $p$ within the coloured region shown in the left panel, and on the bottom nullcline $\eta=0$ as shown in the right panel. Colour denotes the linear growth rate of the unstable modes, with lighter yellow denoting higher growth rate.

\section{Dynamics of interacting waves}
\label{sec: Dynamics of interacting waves}

Even the relatively simple situation of three interacting waves contains a rich diversity of phenomena.  These include periodic trajectories as well as special cases such as the separatrices and interior fixed points. 

\subsection{Periodic solutions}
\label{ssec:periodic solutions}

\begin{figure}
\centering
\includegraphics[width=0.9\linewidth]{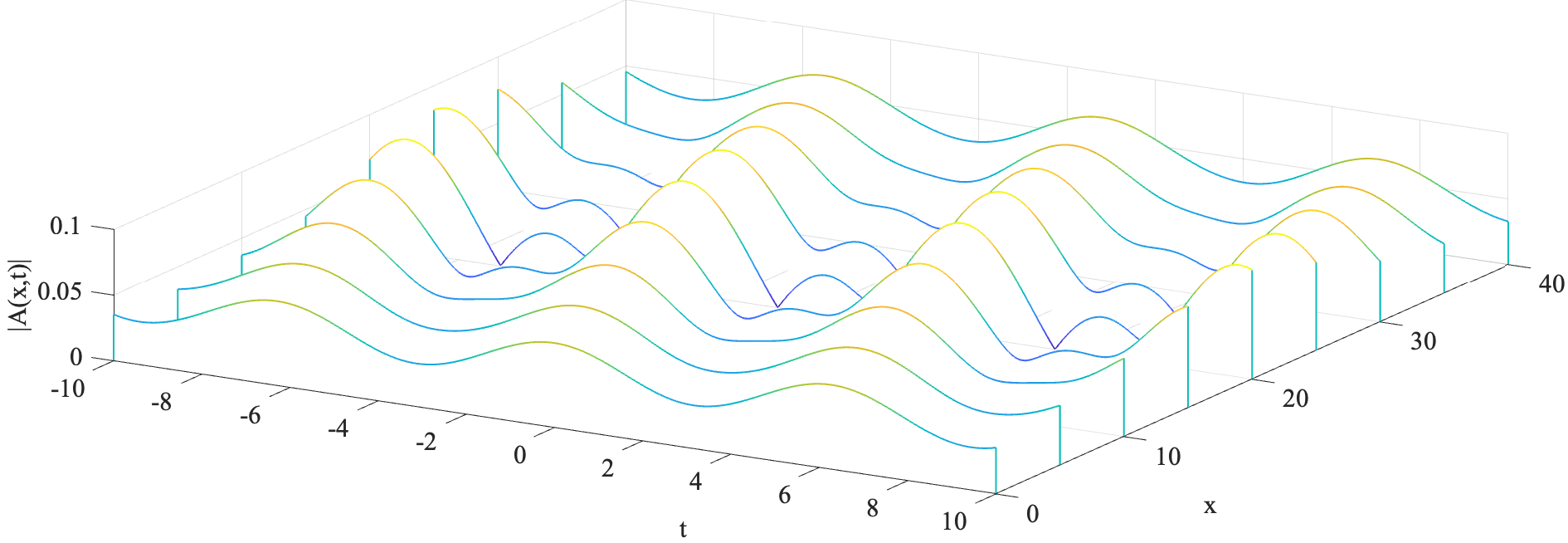}\\
\includegraphics[width=0.45\linewidth]{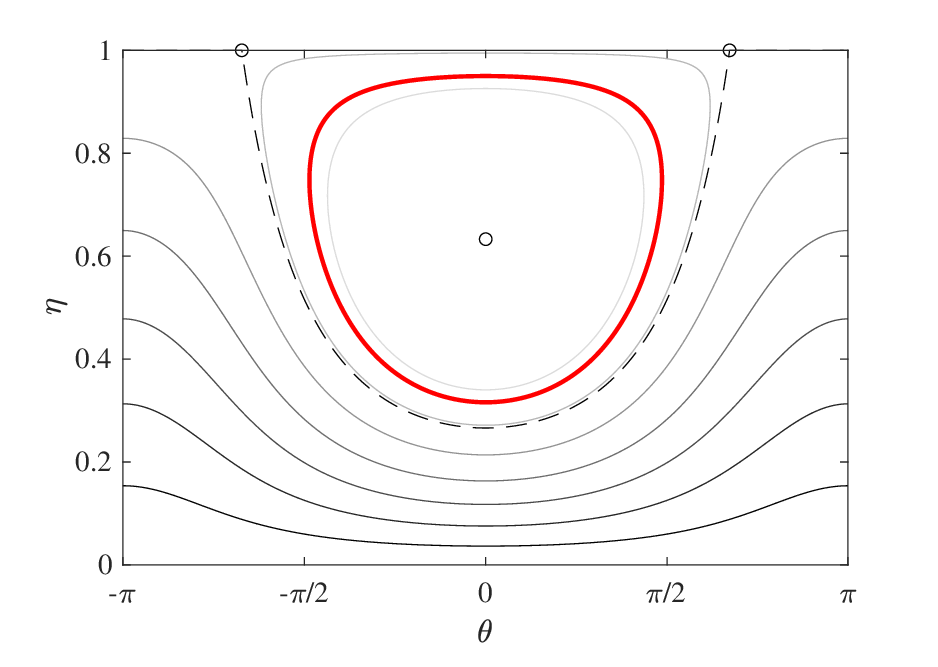}
\includegraphics[width=0.45\linewidth]{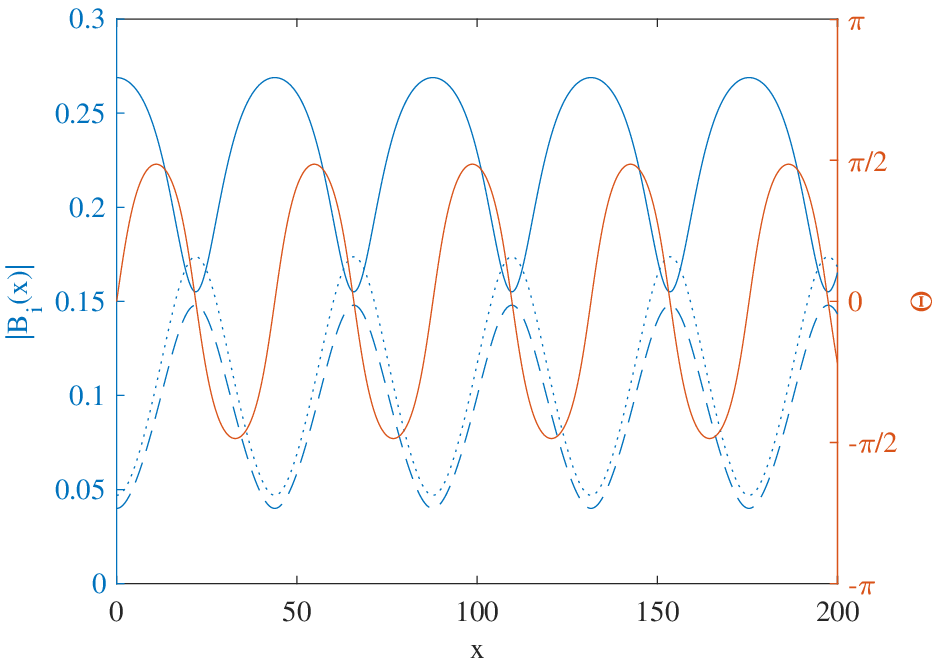}
\caption{Three views of a period solution: (top panel) free-surface envelope in space and time. (Bottom left) solution depicted in phase space (red curve). (Bottom right) Fourier amplitudes $|B_i(x)|$ and dynamic phase $\Theta(x)$ plotted with distance $x$.}
\label{fig:periodic-solution}
\end{figure}

It is well known that the generic, long-time behaviour found when Benjamin-Feir instability is triggered is a periodic recurrence, known as the Fermi-Pasta-Ulam-Tsingou recurrence. Indeed, this periodic behaviour can be readily observed in our solutions. Figure \ref{fig:periodic-solution} shows such a solution from three vantage points. We consider a wave configuration with carrier frequency $f=1$ Hz and carrier steepness $\epsilon=0.2$. This fixes the total energy of the system, but any of the dynamics found in Figure \ref{fig:phase portraits} are possible. We can fix a particular phase portrait by selecting the mode separation $p$, which is chosen to be unity to yield a configuration where the  carrier is unstable to disturbances.

We can select a particular trajectory in phase space by computing the Hamiltonian for given values of $\eta$ and $\Theta.$ In the lower left panel of Figure \ref{fig:periodic-solution} the red curve is the contour for $\eta=0.95$ when $\Theta=0.$ We see that this periodic energy exchange is characterised by a being confined within the separatrix surrounding the centre point $\mathbf{p}_1$ (see \eqref{eq:fp1}), and the dynamic phase $\Theta$ takes on values between approximately $-1.5$ and $1.5$ only. These trajectories are ``unwrapped" in the lower right panel, which shows the individual Fourier amplitudes $|B_i|$ as well as the dynamic phase as functions of $x.$ The interplay between dynamic phase and Fourier amplitudes can be clearly observed: $\Theta(x)$ is at a maximum or minimum when the energy exchange is greatest. Finally the free surface envelope $|A(x,t)|$ is shown for one recurrence period ($x\approx40$ m) in the top panel of the same figure.

\subsection{Breather solutions}
In addition to the recurrent solutions, which are those usually encountered in numerical simulations, there are also solutions with asymptotic behaviour. These solutions tend asymptotically to either the bichromatic or monochromatic wave field, and are therefore termed breathers (in that context we usually speak of a mono/bichromatic \textit{background}).
We can look for breather solutions of our equations by considering the orbits written in the form 
\begin{equation}
    \frac{d \eta}{d \Theta} = \frac{\beta_2 \eta (\eta-1) \sin(\Theta)}{\beta_0 + \beta_1 \eta + \beta_2(2\eta-1) \cos(\Theta)}.
\end{equation}
This can be integrated explicitly, although it is advisable to simplify first. Solutions of interest are those for which $\eta \rightarrow 1,0$ for $\Theta \rightarrow \Theta^*,$ where $\Theta^*$ is the dynamic phase corresponding to a fixed point on the boundary of the phase plane. 

\subsubsection{Breather solutions with monochromatic background}
\label{sssec:Breather monochromatic background}
\begin{figure}[h]
\centering
\includegraphics[width=0.7\linewidth]{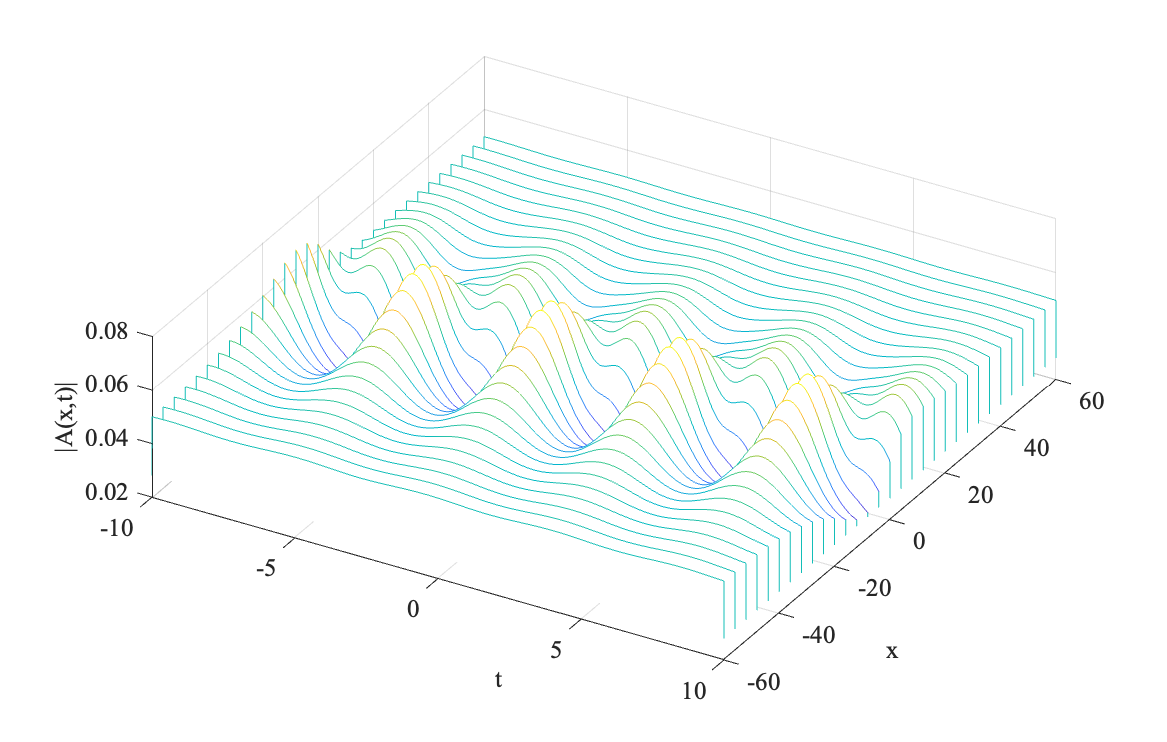}
\caption{The free surface envelope of a breather with monochromatic background, corresponding to $f=1$ Hz, $\epsilon = 0.2,$ and $p=1.4.$ The focusing occurs at $x=0,$ which is the minimum of the energy scale parameter $\eta$ along the separatrix. As $x$ tends to $\pm \infty,$ $\eta$ tends toward 1 and the wave field asymptotically becomes monochromatic.}
\label{fig:breather-envelope-1-1}
\end{figure}

When we are looking for explicit solutions with monochromatic background, i.e.\ separatrices which connect two fixed points at $\eta=1,$ we find the explicit expression
\begin{equation}
    \eta(\Theta) = \frac{\beta_2 \cos(\Theta) - (\beta_2 \cos(\Theta) + \beta_0 + \beta_1) - \beta_0}{\beta_1 + 2 \beta_2 \cos(\Theta)},
\end{equation}
which tends towards 1 as $\Theta \longrightarrow \Theta_0,$ for 
\[ \Theta_0 = \arccos\left( \frac{-\beta_0 - \beta_1}{\beta_2} \right)\]
the dynamic phase of the fixed point.
The differential equation governing $\Theta$ can also be integrated, and the solution written as 
\begin{equation}
    \tan\left(\frac{\Theta}{2}\right) = \textrm{sgn}(\sin(\Theta_0)) \tan\left(\frac{\Theta_0}{2}\right) \tanh\left(\frac{\beta_2}{2}x \sin(\Theta_0)\right).
\end{equation} %\footnote{A solution with $+$ after the first term $\beta_2 cos(\Theta)$ is also valid, but does not yield the correct behaviour as $\Theta=0$.} 
This breather solution corresponds to the class of separatrix found in panel (e) of Figure \ref{fig:phase portraits}. A depiction of the free surface envelope in space and time is shown in Figure \ref{fig:breather-envelope-1-1}. It should be noted that the envelope of a monochromatic wave field is a constant, as observed for large values of $|x|$. At the focusing location $x=0$ the breather is periodic in time, analogous to the well-known Kuznetsov-Ma breather solution of the nonlinear Schr\"odinger equation.

\subsubsection{Breather solutions with bichromatic background}
\label{sssec:Breather bichromatic background}

\begin{figure}[h]
\centering
\includegraphics[width=0.7\linewidth]{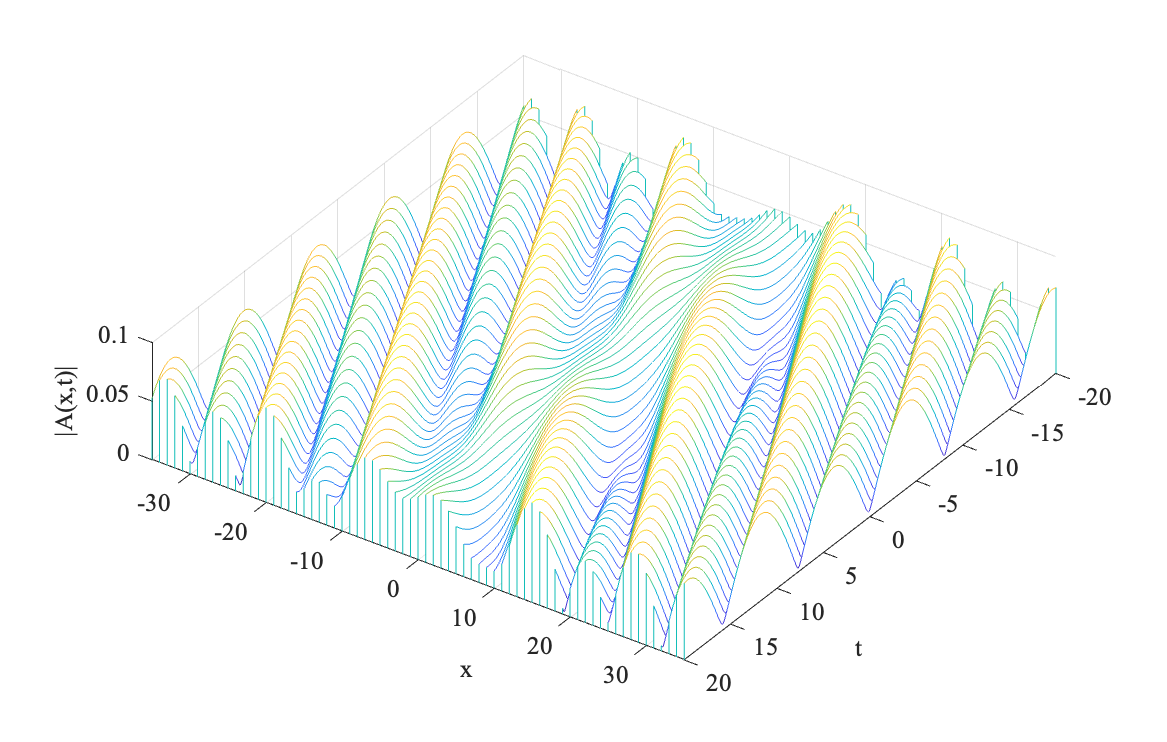}
\caption{The free surface envelope of a breather with bichromatic background, corresponding to $f=1$ Hz, $\epsilon = 0.2,$ and $p=0.4.$ The focusing (in the form of a ``demodulation") occurs at $x=0,$ which is the maximum of the energy scale parameter $\eta$ along the separatrix. At this location $x=0$ the wave field is close to monochromatic, as components $b$ and $c$ are very small. As $x$ tends to $\pm \infty$, $\eta$ tends toward 0 and the wave field asymptotically becomes bichromatic.}
\label{fig:breather-envelope-0-0}
\end{figure}

We may apply the same process to the $\eta=0$ fixed point, which has breathers associated with the separatrices shown in panels (a) and (b) of Figure \ref{fig:phase portraits}. Using $H = 0$ and simplifying equation \eqref{eq:dyn1}, we have an exact solution $\eta(\Theta)$ given by
\begin{align}
    \eta(\Theta) &= \frac{\beta_0 -\beta_2\cos(\Theta)}{\beta_2(1-\cos(\Theta)) - \beta_2-\beta_1/2},
\end{align}
where it can be seen that $\eta\rightarrow 0$ as $\Theta\rightarrow \Theta^*$ where 
\begin{align}
    \Theta^* &= \cos^{-1}\left(\frac{\beta_0}{\beta_2}\right).
\end{align}
We can also solve explicitly the differential equation for $\Theta$ which yields
\begin{align}
    \tan\left(\frac{\Theta}{2}\right) &= \frac{\beta_2}{\beta_2+\beta_0}\sin(\Theta^*)\tanh\left(\frac{\beta_2}{2}\sin(\Theta^*)x\right),
\end{align}
As above, the corresponding free surface envelope is shown in Figure \ref{fig:breather-envelope-0-0}.

\subsubsection{Breather solutions with maximum depletion}
\label{sssec:Breather maximum depletion}

The most striking breather solution occurs when a separatrix connects a fixed point at the top of the phase space $\eta=1$ with a fixed point at the bottom of the phase space $\eta=0$. For this breather solution to exist we require that the fixed points at $\eta=0$ and $\eta=1$ have the same phase, which requires
\begin{equation}
  \beta_0 = -\frac{\beta_1}{2}.
\end{equation}
Such separatrices are vertical orbits $\Theta=\Theta_c$, as shown in Panel (c) of Figure \ref{fig:phase portraits}.
Consequently they are simple to obtain from the dynamical system, since \eqref{eq:dyn1} decouples from \eqref{eq:dyn2}, and can be immediately integrated as a separable ODE for $\eta.$ The solution is
\begin{equation}    \label{eq:1-2 breather sol}
 \eta(x) = \frac{1}{1+C\exp(-\beta_2 x \sin(\Theta_c))}. \end{equation}

For unidirectional waves as considered herein the coefficient $\beta_2$ is strictly non-negative, so that the behaviour of the solution as $x \rightarrow \pm \infty$ depends only on the sign of $\sin(\Theta_c).$ For symmetry, the constant of integration $C$ can be chosen as unity, such that $\eta(0)=1/2$ is centred in the phase space. The explicit formula for the optimal conversion breather  \eqref{eq:1-2 breather sol} also allows us to calculate analytically the evolution length. 

\begin{figure}
\centering
\includegraphics[width=0.7\linewidth]{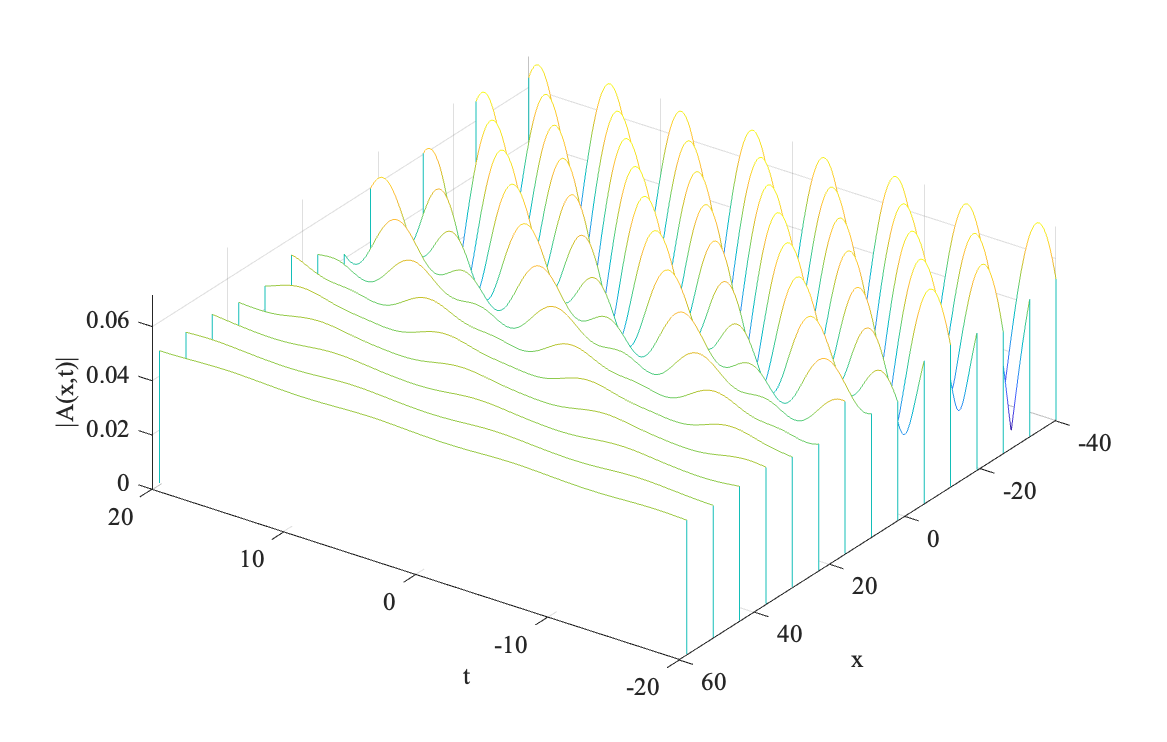}
\caption{The free surface envelope of a breather with maximal depletion, corresponding to $f=1$ Hz, $\epsilon = 0.2,$ and $p\approx  0.5791.$ For $x\rightarrow\infty$ the wave field tends toward a monochromatic state, while for $x\rightarrow-\infty$ the wave tends towards the bichromatic state.}
\label{fig:breather-envelope-0-1}
\end{figure}

The free surface envelope of this breather is shown in Figure \ref{fig:breather-envelope-0-1}. 
This solution is particularly interesting, since it represents the optimal conversion of energy (or transformation) from a monochromatic wave train to a bichromatic wave train (or vice versa). Thus, while any instability of the monochromatic carrier (shown in the left panel of Figure \ref{fig:stability regions top/bottom}) gives rise to a breather of the type discussed in Section \ref{sssec:Breather monochromatic background}, the optimal energy transfer is obtained for a unique value of mode separation for a given carrier steepness -- shown as the red curve in both panels of Figure \ref{fig:stability regions top/bottom}. In addition, the phases of the waves must be tuned in order to obtain this solution, see panel (c) of Figure \ref{fig:phase portraits}. A key observation is that, for a given carrier, the maximal energy transfer occurs at for much closer side-bands (smaller $p$) than the fastest linear growth rate.

\section{Observability of discrete wave interactions}
\label{sec:Observability of discrete wave interactions}

A principal advantage of the spatial Zakharov equation over its better-known temporal sister equation is that it relates directly to properties that can be measured in wave flume experiments. Analogous interaction equations can be used to describe other nonlinear dispersive media, such as electromagnetic Kerr media \cite{Onorato2016b,Xu2020}. A natural question therefore concerns the significance of the three-mode truncation results for experimental work, in particular for optimising energy transfer between a carrier wave and its side bands. 

\begin{figure}
\centering
\includegraphics[width=0.65\linewidth]{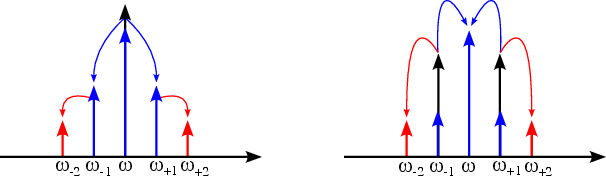}
\caption{Instabilities of monochromatic and bichromatic waves shown embedded in a discrete spectrum, illustrating the possibilities for spectral broadening {through energy transfer}. (Left panel) The Benjamin-Feir instability of a monochromatic wave $f$ (black arrow) transfers energy to $f_{\pm1}$ (blue arrows), but may continue to transfer energy to outlying modes $f_{\pm2}.$ (Right panel) The instability of a bichromatic wave train $f_{\pm1}$ (black arrows) to monochromatic disturbances transfers energy to $f$ (blue arrows), but may also transfer energy to outlying modes $f_{\pm2}$ (red arrows).}
\label{fig:Multi-Mode-Schematic}
\end{figure}

The two interactions that occur in our dynamical systems description of the Benjamin-Feir instability are transfers of energy from one mode $\omega$ to two modes $\omega_{\pm1}$ and vice versa. These are depicted schematically in Figure \ref{fig:Multi-Mode-Schematic}.  The principal energy exchange is indicated by the blue arrows, moving either from one mode to two (left panel) or two modes to one (right panel). What occurs when additional Fourier modes are incorporated? If these are equidistant modes $\omega_{\pm2}, \, \omega_{\pm3}, \ldots$ new energy exchanges become available, and the possibility of spectral broadening appears.  The most important interaction is with the superharmonics $\omega_{+2}$ and $\omega_{-2},$ which is easily subsumed into the foregoing theory.

We will limit our discussion to two cases: the periodic recurrence discussed in Section \ref{ssec:periodic solutions} and the optimal energy conversion from  Section \ref{sssec:Breather maximum depletion}, and investigate whether these are robust from the expanded point of view which allows for higher harmonics. We shall begin with the case of optimal energy conversion, which 
demonstrates a particular instability whereby small perturbations -- with specially tuned mode separation and phases -- are able to asymptotically convert a monochromatic wave train into a bichromatic {state}, or vice versa. 

\begin{figure}
\centering
\includegraphics[width=0.6\linewidth]{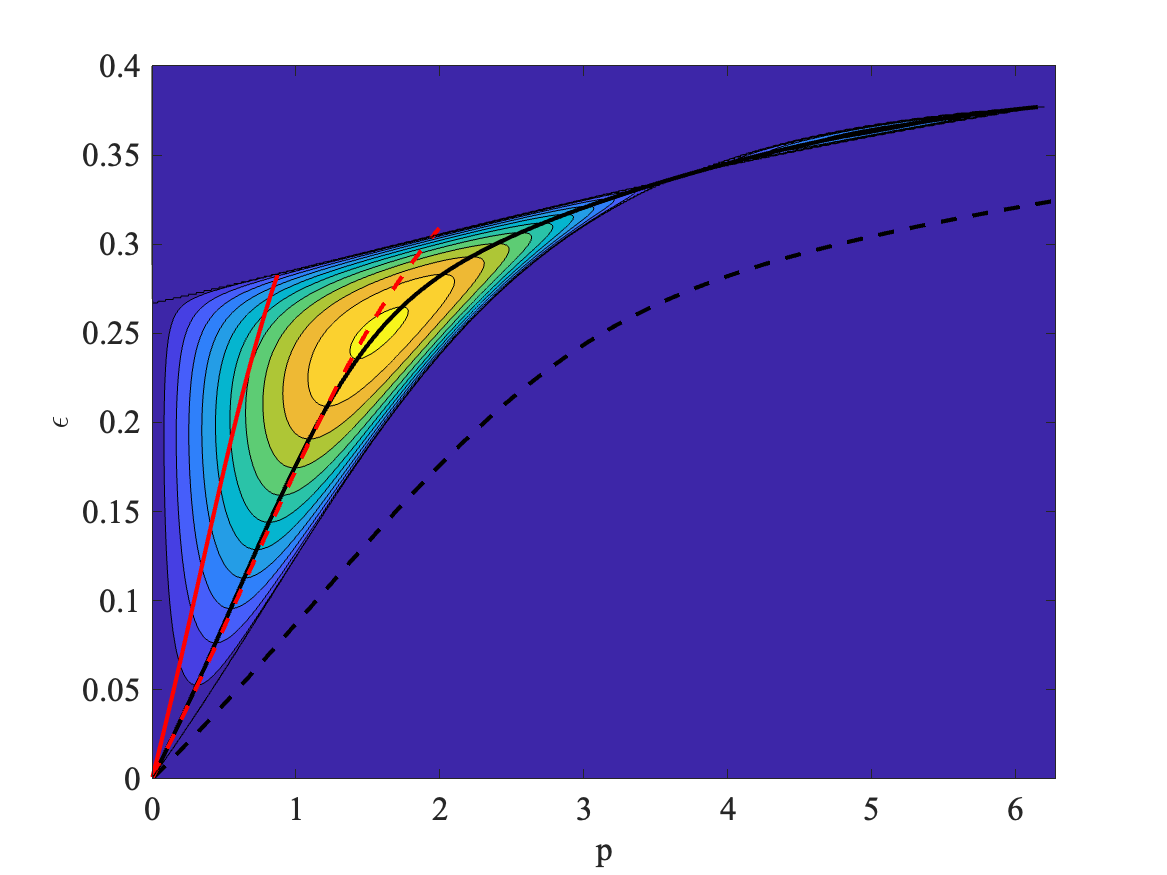}
\caption{Instability domain and linear growth rates (colours) for a monochromatic wave field (as in Figure \ref{fig:stability regions top/bottom}), showing the optimal conversion breather (solid red curve), largest growth rate (black curve), and twice the mode separation of both of the aforementioned (dashed red curve -- breather superharmonic, dashed black curve -- largest growth rate superharmonic).}
\label{fig:instability-domain-supplementary}
\end{figure}

\begin{figure}
\centering
\includegraphics[width=0.95\linewidth]{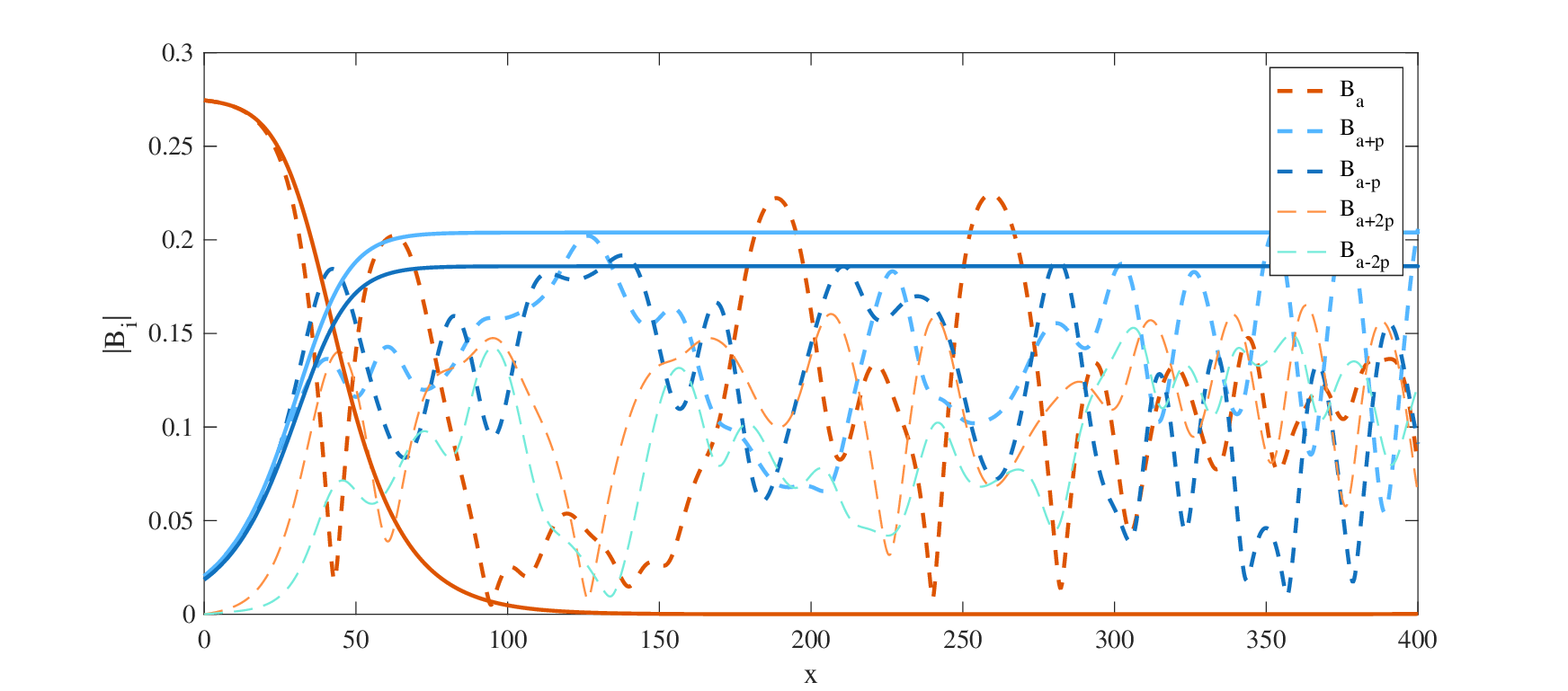}\\
\includegraphics[width=0.95\linewidth]{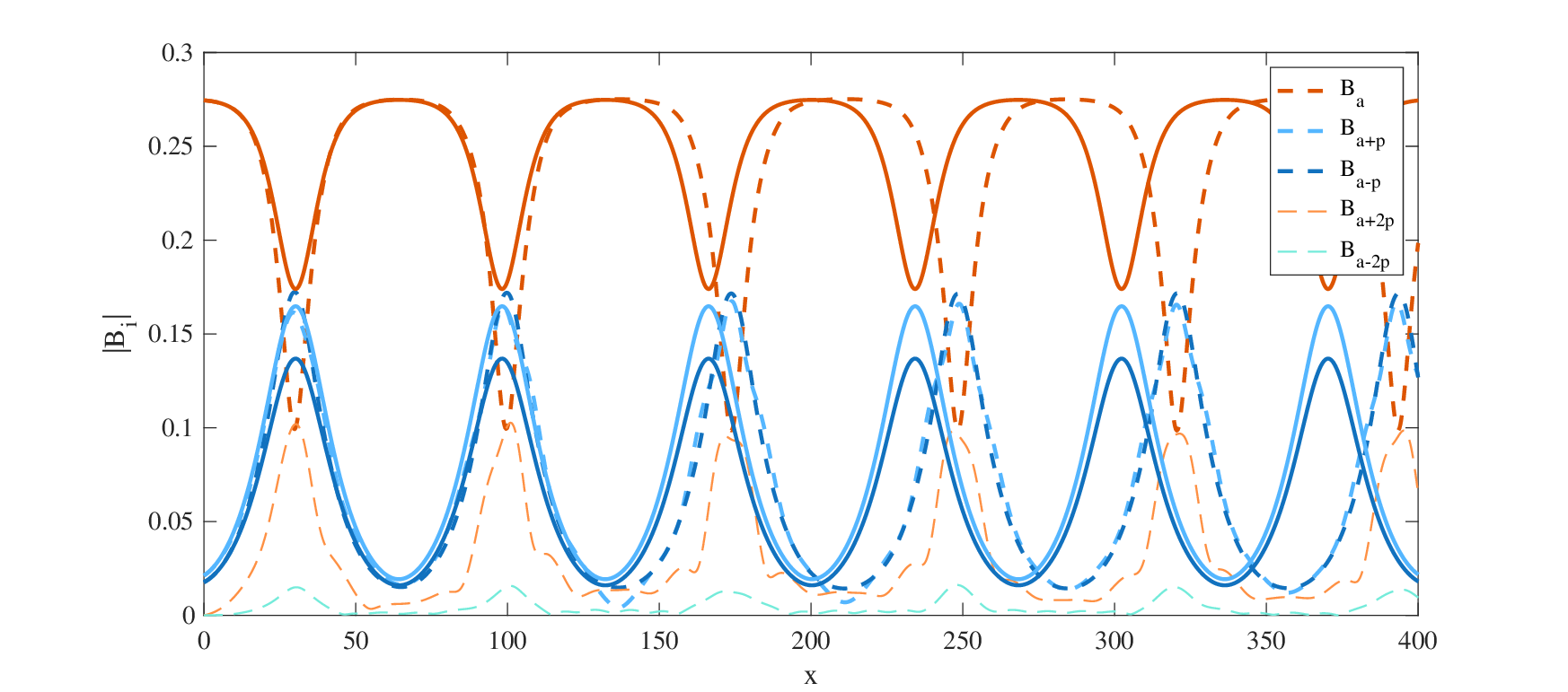}
\caption{Numerical integration with 3 and 5 Fourier modes, with carrier wave $f_a=1$ Hz, $\epsilon_a=0.2.$ (Top panel) Three mode initial conditions corresponding to the maximal energy transfer breather solution found in Section \ref{sssec:Breather maximum depletion}, with $p\approx 0.5798.$ The three-mode system (solid curves) behaves as expected, with energy transferring asymptotically from $B_a$ to $B_{a+p}$ and $B_{a-p}.$ The five-mode system (dashed curves) behaves analogously for very short times, but the fact that modes $f_{a+\pm2p}$ are unstable gives rise to energy exchange and chaotisation of the trajectories. (Bottom panel) Three mode initial conditions corresponding to the largest linear growth rate with $p\approx 1.1532.$ The three mode system (solid curves) undergoes periodic energy exchange. As the higher harmonics are linearly stable these do not participate in any significant energy exchange, but are entrained in the dynamics of the principal triad $a, a+p, a-p.$}
\label{fig:Multi-mode-integration-breather}
\end{figure}

One problem is immediately evident from Figure \ref{fig:phase portraits}: given a carrier steepness and frequency, we find the optimal conversion for a particular value of $p$ which we shall call $p_c$ (panel (c) Figure \ref{fig:phase portraits}). However the same carrier is also unstable to superharmonic perturbations with $2p_c$ (panel (e) Figure \ref{fig:phase portraits}). Moreover, the growth rate of the superharmonics $2p_c$ is larger than that of the fundamental $p_c,$ as can be seen in Figure \ref{fig:instability-domain-supplementary}. In fact, perturbations with $2p_c$ (dashed red curve) are very close to the curve of maximum growth rate (black curve), and so will be preferentially amplified. This is precisely the scenario shown in the top panel of Figure \ref{fig:Multi-mode-integration-breather}, which compares the same initial conditions for a three-mode and a five-mode system.\footnote{Tackling the same problem from the other side of the phase plane, and attempting to make a bichromatic into a monochromatic wave suffers from similar problems. While $\omega_{-1}$ may transfer energy to mode $\omega_0$ in the desired interaction $\omega_{-1} + \omega_{1} = 2 \omega_0$, it much prefers to transfer energy to both $\omega_0$ and $\omega_{-2}$ via the equally accessible interaction $2 \omega_{-1} = \omega_0 + \omega_{-2}.$}

The foregoing discussion highlights some of the differences between a strict truncation and the subsequent evolution when more modes are allowed into the interaction. The analogous arguments can be made for periodic recurrences, which exist alongside the asymptotic solutions and fixed points. In Figure \ref{fig:instability-domain-supplementary} we note that the most unstable perturbation $p_{\text{max}}$ for a given carrier steepness $\epsilon$ (shown in the solid black curve) is not susceptible to superharmonic instabilities since the curve of $2p_{\text{max}}$ lies outside the instability domain. Again, this fact is borne out by numerical simulations comparing the three mode and five mode system, as shown in the bottom panel of Figure \ref{fig:Multi-mode-integration-breather}.

It may seem initially surprising that the superharmonics $\omega_{\pm 2p}$ oscillate despite being linearly stable. This can be seen from linear stability analysis (see \ref{ssec:Linear stability analysis}), which shows that higher harmonics are entrained and grow linearly in concert with the dominant unstable triad $\omega, \, \omega_{\pm1}$. While the qualitative behaviour, and in particular the recurrent nature of the solution, is preserved some quantitative differences do manifest, among these the aperiodicity of the five-mode solution. Another significant difference is the extent to which the carrier (denoted $|B_a|$) can deplete: in a case with more modes the cascade of energy from the carrier into the sidebands and superharmonics leads to a greater depletion of the carrier amplitude itself -- a process that is brought to its apotheosis by Akhmediev or Kuznetsov-Ma breathers \cite{Kuznetsov1977,Akhmediev1985} (see Chin et al \cite{Chin2015}).

\section{Discussion}
\label{sec: Discussion}

We have set out to examine the spatial Benjamin-Feir instability from the perspective of the spatial Zakharov equation. While the temporal case has been studied very extensively, this physically important case has received much less attention. In order to employ techniques of phase plane analysis we restrict ourselves to the three Fourier modes forming a degenerate quartet, the germ of the Benjamin-Feir instability. With this restriction it is possible to describe the entire subsequent behaviour simply and analytically -- classifying the phase portraits, identifying separatrices with breathers and fixed points with steady-state solutions, see \cite{Liao2016}.

Two particularly important insights are obvious from our approach, but obscured by the classical treatment of linear stability analysis and subsequent numerical integration. The first is that the largest (linear) growth rate does not correspond to the largest energy exchange among the modes. This is clear from Figure \ref{fig:stability regions top/bottom}, where the latter is shown in the red curves, and is observed to occur consistently for a smaller value of mode separation than the highest growth rate at equal carrier steepness. Just because the a disturbance grows the fastest does not mean that it grows the most. 

A second key insight concerns the stability thresholds. Monochromatic and bichromatic waves are identified with the top and bottom nullclines of our dynamical system, respectively, and their orbital stability is determined by the existence of fixed points thereon. Indeed the existence criteria for such fixed points are identical to the linear stability thresholds. However, the phase portraits (see Figure \ref{fig:phase portraits}, panel (f)) make clear that energy exchange persists outside this instability threshold, as $\eta$ changes along the trajectories. Contrary to what one might expect from the linear analysis alone, even stable configurations generally exhibit oscillations between the Fourier amplitudes.

Breathers are famed exact solutions of the nonlinear Schr\"odinger equation \cite{Akhmediev1987}, and can be observed experimentally \cite{chabchoub2011rogue,Chabchoub2012,Kibler2010}, which raises the question whether they can appear in more general equations governing inviscid propagation of water waves. Early numerical simulations affirming this using the equations for potential flow were performed by Dyachenko \& Zakharov \cite{Dyachenko2008}. A more detailed study was undertaken by Slunyaev \& Shrira \cite{Slunyaev2013} using numerical solutions of the Euler equation, who found that breathers carefully initialised with NLS initial conditions could be propagated numerically without significant change. This is quite surprising given the simple nature of the NLS and the numerous restrictions made in its derivation.

A natural question is whether breather solutions exist within the framework of the reduced Zakharov equation, which -- while limited to third order in nonlinearity -- is at least free of the bandwidth restrictions which appear in the NLS. The answer to this question is positive, although no explicit  expressions comparable to those found by Akhmediev et al. \cite{Akhmediev1987} have been found. Using the Petviashvili method, pioneered in investigations of the so-called compact Dyachenko-Zakharov equation by Fedele \& Dutykh \cite{Fedele2012}, Kachulin et al \cite{Kachulin2019} have successfully obtained a numerical breather solution of the spatial Zakharov equation. 

As is the case for the Akhmediev breather solution of the NLS \cite{Chin2015}, this solution manifests the Benjamin-Feir instability and so must occur within the linear instability domain (see Figure \ref{fig:stability regions top/bottom}). However, its higher harmonics must be linearly stable, in order that the energy transfer is reversible and the monochromatic background is obtained as $x\rightarrow \pm \infty.$ In analogy with the situation found in the NLS by Bendahmane et al. \cite{Bendahmane2015}, it is likely that the largest realisable depletion of the carrier wave is to be found along this breather trajectory.

Finding an explicit Kuznetsov-Ma type breather solution of the Zakharov equation is not so straightforward due to the high-dimensional phase-space involved. Such a solution, tending asymptotically to a fixed point on the submanifold consisting of monochromatic waves would be a kind of coda to Section \ref{sec:Observability of discrete wave interactions}, presenting an exactly reversible cascade of energy to the superharmonics. Evidently, finding an experimental setting where the distinct features of such a breather could be realised would also be of great interest.

\appendix 
 
\section{Linear stability analysis for the spatial Zakharov equation}
\label{ssec:Linear stability analysis}

It may be surprising to find that, even when the modes $\omega_{\pm2}$ are linearly stable, they nevertheless grow from zero. We can appreciate this fact by a closer look at the linear stability analysis of the  discretised spatial Zakharov equation \eqref{eq:Spatial ZE Discr}
\begin{equation}
    i c_{g,j} b_j'(x) = - \frac{\omega_j}{2} b_j + \sum T_{jlmn} b_l^* b_m b_n \delta_{jlmn}.
\end{equation}
Note that $k_j c_{g,j} = \omega_j/2.$

If we initially assume that mode $b_0 \gg b_i$ for all $i \neq 0,$ and we neglect products of small terms, we obtain the single equation 
\begin{equation}
    i c_{g,0} b_0' = - \frac{\omega_0}{2}b_0 + T_{0,0} |b_0|^2 b_0 = \Omega_0 b_0.
\end{equation}
where we write $\Omega_i = -\frac{\omega_i}{2} + T_{0,i} |b_0|^2.$ The solution to this equation is the Stokes' wave (see Section \ref{sec:Simple Solutions}) 
\[ b_0 = A_0 \exp(-i \Omega_0/c_{g,0}). \]

Neglecting the second harmonics $b_{\pm 2}$ as small, and retaining only terms linear in $b_{\pm1}$ we obtain the system of equations
\begin{align}
    i c_{g,1} b_1' &= \Omega_1 b_1 + T_{-1,1,0,0} b_{-1}^* b_0^2, \\
    i c_{g,-1} b_{-1}' &= \Omega_{-1} b_{-1} + T_{-1,1,0,0} b_{1}^* b_0^2.
\end{align}
This is a linear system in the side-bands $b_{\pm1}$, and can be solved by substituting the Ansatz
\begin{align*}
&b_1 = A_1 \exp\left(x\left(\sigma - i \frac{\Omega_1}{c_{g,1}} - i \frac{\beta}{2}\right)\right),\\
&b_{-1} = A_{-1} \exp\left(x\left(\sigma^* - i \frac{\Omega_{-1}}{c_{g,{-1}}} - i \frac{\beta}{2}\right)\right),
\end{align*}
where $\beta=-\beta_0-\beta_1 = 2\Omega_0/c_{g,0} - \Omega_1/c_{g,1} - \Omega_{-1}/c_{g,-1}.$ Substitution shows that this linear system has a solution when the determinant of the coefficient matrix vanishes, precisely the condition \eqref{eq:linstab} previously obtained from phase-plane analysis.

At the next order, assuming $b_{\pm2}$ are small and retaining terms containing $b_0, \, b_{\pm 1},$ one obtains the linear system
\begin{align} 
i c_{g,2} \frac{d b_2}{dt} &= 2 T_{-1,2,0,1} b_{-1}^* b_0 b_1 + T_{0,2,1,1} b_0^* b_1^2, \\ 
i c_{g,-2} \frac{d b_{-2}}{dt} &= 2 T_{-2,1,-1,0} b_1^* b_{-1} b_0 + T_{-2,0,-1,-1} b_0^* b_{-1}^2.
\end{align}
These are forced equations for modes $\pm2,$ which show that these modes grow linearly due to the dominant interaction between the carrier and the side-bands $\pm1.$

\section*{Funding sources}
RS and CH acknowledge the support of EPSRC Grant EP/V012770/1.


\begin{thebibliography}{10}

\bibitem{Ablowitz1990}
M.~J. Ablowitz and B.~M. Herbst.
\newblock {On homoclinic structure and numerically induced chaos for the
  nonlinear Schrodinger equation}.
\newblock {\em SIAM J. Appl. Math.}, 50(2):339--351, 1990.

\bibitem{Akhmediev1985}
N.~Akhmediev, V.~Eleonskii, and N.~Kulagin.
\newblock Generation of periodic trains of picosecond pulses in an optical
  fiber: exact solutions.
\newblock {\em Sov. Phys. JETP}, 62(5):894--899, 1985.

\bibitem{Akhmediev1987}
N.~N. Akhmediev, V.~M. Eleonski\u{\i}, and N.~E. Kulagin.
\newblock First-order exact solutions of the nonlinear {S}chr\"{o}dinger
  equation.
\newblock {\em Teoret. Mat. Fiz.}, 72(2):183--196, 1987.

\bibitem{Andrade2023instability}
D.~Andrade and R.~Stuhlmeier.
\newblock Instability of waves in deep water---a discrete hamiltonian approach.
\newblock {\em European Journal of Mechanics-B/Fluids}, 101:320--336, 2023.

\bibitem{Andrade2023}
D.~Andrade and R.~Stuhlmeier.
\newblock {The nonlinear Benjamin-Feir instability - Hamiltonian dynamics,
  discrete breathers, and steady solutions}.
\newblock {\em Journal of Fluid Mechanics}, 958:A17, 2023.
\newblock doi:10.1017/jfm.2023.96.

\bibitem{Bendahmane2015}
A.~Bendahmane, A.~Mussot, A.~Kudlinski, P.~Szriftgiser, M.~Conforti,
  S.~Wabnitz, and S.~Trillo.
\newblock {Optimal frequency conversion in the nonlinear stage of modulation
  instability}.
\newblock {\em Opt. Express}, 23(24):30861, 2015.

\bibitem{Benjamin1967a}
T.~B. Benjamin and J.~E. Feir.
\newblock {The disintegration of wave trains on deep water Part 1. Theory}.
\newblock {\em J. Fluid Mech.}, 27(03):417--430, 1967.

\bibitem{Benney1967}
D.~Benney and A.~Newell.
\newblock The propagation of nonlinear wave envelopes.
\newblock {\em Journal of mathematics and Physics}, 46(1-4):133--139, 1967.

\bibitem{Bespalov1966}
V.~I. Bespalov and V.~I. Talanov.
\newblock Filamentary structure of light beams in nonlinear liquids.
\newblock {\em Soviet Journal of Experimental and Theoretical Physics Letters},
  3:307, 1966.

\bibitem{Bogoliubov1947}
N.~Bogoliubov.
\newblock On the theory of superfluidity.
\newblock {\em J. Phys}, 11(1):23, 1947.

\bibitem{Bretherton1964}
F.~Bretherton.
\newblock {Low frequency oscillations trapped near the equator}.
\newblock {\em Tellus}, 16(2):181--185, 1964.


\bibitem{Bustamante2009a}
M.~D. Bustamante and E.~Kartashova.
\newblock {Effect of the dynamical phases on the nonlinear amplitudes'
  evolution}.
\newblock {\em Europhys. Lett.}, 85(3), 2009.

\bibitem{Cappellini1991}
G.~Cappellini and S.~Trillo.
\newblock {Third-order three-wave mixing in single-mode fibers: exact solutions
  and spatial instability effects}.
\newblock {\em J. Opt. Soc. Am. B}, 8(4):824, 1991.

\bibitem{Chabchoub2012}
A.~Chabchoub, N.~Akhmediev, and N.~P. Hoffmann.
\newblock {Experimental study of spatiotemporally localized surface gravity
  water waves}.
\newblock {\em Phys. Rev. E}, 86(1):016311, jul 2012.

\bibitem{Chabchoub2016a}
A.~Chabchoub and R.~H. Grimshaw.
\newblock {The hydrodynamic nonlinear Schr{\"{o}}dinger equation: Space and
  time}.
\newblock {\em Fluids}, 1(3):1--10, 2016.

\bibitem{chabchoub2011rogue}
A.~Chabchoub, N.~Hoffmann, and N.~Akhmediev.
\newblock Rogue wave observation in a water wave tank.
\newblock {\em Physical Review Letters}, 106(20):204502, 2011.

\bibitem{Chin2015}
S.~A. Chin, O.~A. Ashour, and M.~R. Beli{\'{c}}.
\newblock {Anatomy of the Akhmediev breather: Cascading instability, first
  formation time, and Fermi-Pasta-Ulam recurrence}.
\newblock {\em Phys. Rev. E - Stat. Nonlinear, Soft Matter Phys.}, 92(6):1--9,
  2015.

\bibitem{Craik1986a}
A.~D.~D. Craik.
\newblock {\em {Wave Interactions and Fluid Flows}}.
\newblock Cambridge University Press, jan 1986.

\bibitem{Crawford1981}
D.~R. Crawford, B.~M. Lake, P.~G. Saffman, and H.~C. Yuen.
\newblock {Stability of weakly nonlinear deep-water waves in two and three
  dimensions}.
\newblock {\em J. Fluid Mech.}, 105:177--191, 1981.

\bibitem{Dyachenko2008}
A.~I. Dyachenko and V.~E. Zakharov.
\newblock On the formation of freak waves on the surface of deep water.
\newblock {\em JETP letters}, 88:307--311, 2008.

\bibitem{Dysthe1979}
K.~B. Dysthe.
\newblock {Note on a modification to the nonlinear Schrodinger equation for
  application to deep water waves}.
\newblock {\em Proc. R. Soc. A Math. Phys. Eng. Sci.}, 369:105--114, aug 1979.

\bibitem{Fedele2012}
F.~Fedele and D.~Dutykh.
\newblock Special solutions to a compact equation for deep-water gravity waves.
\newblock {\em Journal of Fluid Mechanics}, 712:646--660, 2012.

\bibitem{Galvagno2021}
M.~Galvagno, D.~Eeltink, and R.~Stuhlmeier.
\newblock {Spatial deterministic wave forecasting for nonlinear sea-states}.
\newblock {\em Phys. Fluids}, 33(10), 2021.

\bibitem{Kachulin2019}
D.~Kachulin, A.~Dyachenko, and A.~Gelash.
\newblock {Interactions of coherent structures on the surface of deep water}.
\newblock {\em Fluids}, 4(2):1--21, 2019.

\bibitem{Kibler2010}
B.~Kibler, J.~Fatome, C.~Finot, G.~Millot, F.~Dias, G.~Genty, N.~Akhmediev, and
  J.~M. Dudley.
\newblock {The Peregrine soliton in nonlinear fibre optics}.
\newblock {\em Nat. Phys.}, 6(10):790--795, aug 2010.

\bibitem{Krasitskii1994}
V.~P. Krasitskii.
\newblock {On reduced equations in the Hamiltonian theory of weakly nonlinear
  surface waves}.
\newblock {\em J. Fluid Mech.}, 272:1--20, 1994.

\bibitem{Kuznetsov1977}
E.~A. Kuznetsov.
\newblock Solitons in a parametrically unstable plasma.
\newblock 236:575--577, 1977.

\bibitem{Liao2016}
S.~Liao, D.~Xu, and M.~Stiassnie.
\newblock {On the steady-state nearly resonant waves}.
\newblock {\em J. Fluid Mech.}, 794:175--199, 2016.

\bibitem{Longuet-Higgins1962d}
M.~S. Longuet-Higgins and O.~M. Phillips.
\newblock {Phase velocity effects in tertiary wave interactions}.
\newblock {\em J. Fluid Mech.}, 12(3):333--336, 1962.

\bibitem{Ma1979}
Y.-C. Ma.
\newblock The perturbed plane-wave solutions of the cubic schr{\"o}dinger
  equation.
\newblock {\em Studies in Applied Mathematics}, 60(1):43--58, 1979.

\bibitem{Onorato2016b}
M.~Onorato, F.~Baronio, M.~Conforti, A.~Chabchoub, P.~Suret, and S.~Randoux.
\newblock Hydrodynamic and optical waves: A common approach for unidimensional
  propagation.
\newblock {\em Rogue and Shock Waves in Nonlinear Dispersive Media}, pages
  1--22, 2016.

\bibitem{Onorato2016}
M.~Onorato, S.~Residori, and F.~Baronio, editors.
\newblock {\em {Rogue and Shock Waves in Nonlinear Dispersive Media}}.
\newblock Springer, 2016.

\bibitem{Peregrine1983a}
D.~H. Peregrine.
\newblock {Water waves, nonlinear Schr{\"{o}}dinger equations and their
  solutions}.
\newblock {\em J. Aust. Math. Soc. Ser. B. Appl. Math.}, 25(1):16--43, 1983.

\bibitem{Phillips1960}
O.~M. Phillips.
\newblock {On the dynamics of unsteady gravity waves of finite amplitude Part
  1. The elementary interactions}.
\newblock {\em J. Fluid Mech.}, 9(2):193--217, oct 1960.

\bibitem{Shemer2017}
L.~Shemer and A.~Chernyshova.
\newblock {Spatial evolution of an initially narrow-banded wave train}.
\newblock {\em J. Ocean Eng. Mar. Energy}, 3(4):333--351, 2017.

\bibitem{Shemer2001}
L.~Shemer, H.~Jiao, E.~Kit, and Y.~Agnon.
\newblock {Evolution of a nonlinear wave field along a tank: experiments and
  numerical simulations based on the spatial Zakharov equation}.
\newblock {\em J. Fluid Mech.}, 427:107--129, 2001.

\bibitem{Shemer2002}
L.~Shemer, E.~Kit, and H.~Jiao.
\newblock {An experimental and numerical study of the spatial evolution of
  unidirectional nonlinear water-wave groups}.
\newblock {\em Phys. Fluids}, 14(10):3380--3390, oct 2002.

\bibitem{Slunyaev2013}
A.~V. Slunyaev and V.~I. Shrira.
\newblock {On the highest non-breaking wave in a group: Fully nonlinear water
  wave breathers versus weakly nonlinear theory}.
\newblock {\em J. Fluid Mech.}, 735:203--248, 2013.

\bibitem{Stuhlmeier2024}
R.~Stuhlmeier.
\newblock An introduction to the Zakharov equation for modelling deep water
  waves.
\newblock In D.~Henry, editor, {\em Nonlinear Dispersive Waves}. Springer,
  to appear.

\bibitem{trillo1991dynamics}
S.~Trillo and S.~Wabnitz.
\newblock Dynamics of the nonlinear modulational instability in optical fibers.
\newblock {\em Optics letters}, 16(13):986--988, 1991.

\bibitem{TRULSEN1996}
K.~Trulsen and K.~B. Dysthe.
\newblock A modified nonlinear schr{\"o}dinger equation for broader bandwidth
  gravity waves on deep water.
\newblock {\em Wave Motion}, 24(3):281--289, 1996.

\bibitem{Xu2020}
G.~Xu, A.~Chabchoub, D.~E. Pelinovsky, and B.~Kibler.
\newblock Observation of modulation instability and rogue breathers on
  stationary periodic waves.
\newblock {\em Physical Review Research}, 2(3):033528, 2020.

\bibitem{Yuen1982}
H.~C. Yuen and B.~M. Lake.
\newblock {Nonlinear Dynamics of Deep-Water Gravity Waves}.
\newblock In {\em Adv. Appl. Mech.}, pages 68--229. Academic Press, 1982.

\bibitem{Zakharov1968}
V.~Zakharov.
\newblock {Stability of periodic waves of finite amplitude on the surface of a
  deep fluid}.
\newblock {\em J. Appl. Mech. Tech. Phys.}, 9(2):190--194, 1968.

\bibitem{Zakharov1992a}
V.~E. Zakharov, V.~S. L'vov, and G.~Falkovich.
\newblock {\em {Kolmogorov Spectra of Turbulence I}}.
\newblock Springer-Verlag Berlin Heidelberg, 1992.

\end{thebibliography}
\end{document}